\documentclass[11pt]{article}
\pdfoutput=1
\usepackage{jheppubmod}
\usepackage[punctsep]{collref}
\usepackage{comment}
\collectsep[]{;}
\usepackage{hyperref}
\usepackage{subcaption}
\usepackage{feynmp}
\usepackage{color}
\usepackage{latexsym}
\usepackage{bbold}
\usepackage{bm}
\usepackage{slashed}
\DeclareMathOperator*{\SumInt}{%
	\mathchoice%
	{\ooalign{$\displaystyle\sum$\cr\hidewidth$\displaystyle\int$\hidewidth\cr}}
	{\ooalign{\raisebox{.14\height}{\scalebox{.7}{$\textstyle\sum$}}\cr\hidewidth$\textstyle\int$\hidewidth\cr}}
	{\ooalign{\raisebox{.2\height}{\scalebox{.6}{$\scriptstyle\sum$}}\cr$\scriptstyle\int$\cr}}
	{\ooalign{\raisebox{.2\height}{\scalebox{.6}{$\scriptstyle\sum$}}\cr$\scriptstyle\int$\cr}}
}

\def\x{\vec{x}}
\def\y{\vec{y}}

\def\p{\vec{p}}
 \def\dm{\mathrm{d}}
\def\K{\vec{K}}
\def\0{\vec{0}}

\def\q{{\vec{q}}}

\def \k {{\vec{k}}}
\def\Or[#1]{{\text{O}}\left({#1}\right)}
\def\dotl[#1,#2]{\left\langle #1,\, #2 \right\rangle}
\def\dotlb[#1,#2]{\left\langle #1,\, #2 \right\rangle}
\def\dotlm[#1,#2]{\left[ #1,\, #2 \right]}
\def\dotp[#1,#2]{(\vect{#1} \cdot\vect{#2})}
\def\aff[#1,#2]{\hat{#1}(#2)}
\def\n4sym{{\cal N}=4 SYM}
\def\>{\rangle}
\def\<{\langle}
\def\({\left(}
\def\){\right)}
\def\weight[#1,#2,#3]{\{(#1),#2,#3\}}
\def\ads[#1]{$\text{AdS}_{#1}$}

\newcommand{\nn}{\\ \nonumber}

\newcommand{\be}{\begin{equation}}
\newcommand{\ee}{\end{equation}}
\newcommand{\beq}{\begin{eqnarray}}
\newcommand{\eeq}{\end{eqnarray}}
\newcommand{\ba}{\begin{align}}
\newcommand{\ea}{\end{align}}

\renewcommand{\vec}[1]{\boldsymbol{#1}}
\newcommand{\bs}{\begin{split}}
	\def\sess\end{split}

\newcommand{\vect}[1]{{\boldsymbol{#1}}}

\usepackage{setspace} 
\title{Thermalization and chaos in QED$_{3}$}
\author[1]{Julia Steinberg,}
\affiliation[1]{Department of Physics, Harvard University, Cambridge MA
	02138, USA.}
{\tiny }
\author[2]{Brian Swingle}
\affiliation[2]{Condensed Matter Theory Center, Maryland Center for Fundamental Physics, Joint Center for Quantum Information and Computer Science,
and Department of Physics, University of Maryland, College Park, MD 20742, USA}
\date{\today \\
	\vspace{1.6in}}
\begin{document}
	\abstract{We study the real time dynamics of $N_F$ flavors of fermions coupled to a $U(1)$ gauge field in $2+1$ dimensions to leading order in a $1/N_F$ expansion. For large enough $N_{F}$, this is an interacting conformal field theory and describes the low energy properties of the Dirac spin liquid. We focus on thermalization and the onset of many-body quantum chaos which can be diagnosed from the growth of initally anti-commuting fermion field operators. We compute such anti-commutators in this gauge theory to leading order in $1/N_F$. We find that the anti-commutator grows exponentially in time and compute the quantum Lyapunov exponent. We briefly comment on chaos, locality, and gauge invariance.}.
	\maketitle
 \section{Introduction}
Remarkably, closed quantum systems exhibiting chaotic dynamics can act as their own heat bath, leading to an effective thermal state at late time~\cite{Deutsch1991}\cite{Srednicki1994}\cite{Tasaki1998a}\cite{Rigol2008}. One of the great challenges in condensed matter physics is to understand this process of thermalization in closed many-body systems in the vicinity of interacting quantum critical points. These systems have fast dynamics and relax on a timescale set by the temperature, $\tau_{\varphi}\sim \frac{\hbar}{k_B T}$. At early times, the timescales of thermalization can be probed via the time dependence of simple local correlation functions. At intermediate times, a process called ``scrambling" spreads information about the initial state or perturbation throughout the entire system~\cite{Shenker2014}\cite{Hayden2007}\cite{Sekino2008}\cite{Hosur2016}. Scrambling after local relaxation has occurred is not easily accessible in local correlation functions, as it must be described by objects that probe degrees of freedom of the entire system.

In quantum systems with many degrees of freedom, the rate of scrambling can be characterized by a quantum Lyapunov exponent which controls the onset of quantum chaos \cite{larkin1969quasiclassical}\cite{Shenker2014}\cite{Kitaev_talk}. In classical systems, one aspect of chaos is the divergence of initially nearby phase space trajectories. This gives a schematic definition for a classical Lyapunov exponent $\lambda_{L}$ as
\begin{eqnarray}
\vert\delta\x(t)\vert=e^{\lambda_{L}t}\vert\delta\x(0)\vert.
\end{eqnarray}
An analogous notion in quantum systems is obtained from the behavior of an out-of-time-order correlation function (OTOC) \cite{larkin1969quasiclassical}\cite{Shenker2014}\cite{Kitaev_talk}. This object has an elegant description in AdS/CFT as the correlation between two entangled black holes connected by a wormhole \cite{Shenker2014}\cite{Roberts2014}. In fermionic systems, this object is contained in the regulated squared anti-commutator of two fermionic operators. We denote this object as $\mathcal{F}(t,\x)$ which is defined as follows
\begin{equation}
\mathcal{F}(t,\x)=\langle\lbrace W(t,\x),V(0)\rbrace\lbrace W(t),V(0)\rbrace^{\dagger}\rangle_{\beta}
\label{eq:squareanticom}
\end{equation}
where $V$ and $W$ are two fermionic operators separated in space-time.

The squared anti-commutator can be understood as characterizing the effect that one local measurements has on another. For spatially local systems, causality implies that initially separated $V$ and $W$ must have zero anti-commutator (since we are dealing with fermions). However, the squared anti-commutator grows exponentially with time in systems with many local degrees of freedom, and it saturates to its late time value after the ``scrambling" time $t_{\ast}$. During this process, the anti-commutator behaves like
\begin{equation}
\mathcal{F}(t,\x)\sim \epsilon e^{\lambda_{L}\left(t-\frac{|\x|}{v_{B}}\right)}
\end{equation}
where $\lambda_{L}$ is identified as a Lyapunov exponent in analogy with classical chaos, and $v_{B}$ is called the butterfly velocity which acts as an effective speed limit. We emphasize that this is expected to hold when there are a large number of local degrees of freedom. More generally, it has been argued recently that the exponential form above is modified by quantum fluctuations in a non-trivial way. Here we focus for simplicity on the spatially averaged object, $\int d^2 x \mathcal{F}(\x,t)$, which diagnoses scrambling within the local Hilbert space.

Quantum Lyapunov exponents have by now received intense scrutiny. They can be measured experimentally~\cite{swingle2016} \cite{Zhu2016} \cite{Yao2016a} \cite{Halpern2016} \cite{Halpern2017} \cite{Campisi2017} \cite{Yoshida2017} \cite{Garttner2016} \cite{Wei2016} \cite{Li:2017pbq} \cite{Meier2017}, are related to operator growth~\cite{Nahum2017a} \cite{VonKeyserlingk2017} \cite{Xu2018} \cite{Roberts2018} \cite{Jonay2018} \cite{Mezei2018} \cite{You2018} \cite{Chen2018}, and can be computed either numerically or analytically in many model systems~\cite{Kitaev_talk} \cite{Nahum2017a} \cite{VonKeyserlingk2017} \cite{Xu2018} \cite{Sachdev1993} \cite{Gu2017} \cite{Luitz2017} \cite{Bohrdt2017a} \cite{Heyl2018} \cite{Lin2018} \cite{Nahum2017} \cite{Rakovszky2017} \cite{Khemani2017} \cite{Khemani2018} \cite{Lashkari2012}\cite{Shenker2014a} \cite{Aleiner2016} \cite{Swingle:2016jdj} \cite{Grozdanov:2018atb} \cite{Patel2017} \cite{Swingle2018Resilience} \cite{Dressel2018SCweak} \cite{Menezes2018Sonic} \cite{Scaffidi2017}.

	\begin{figure}
    \centering
    \begin{subfigure}[b]{0.4\textwidth}
        \includegraphics[width=\textwidth]{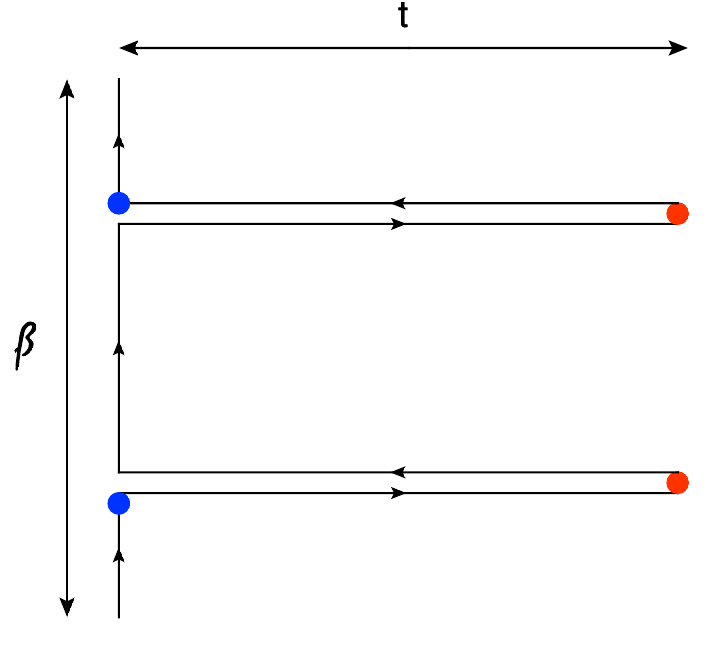}
        \caption{}
        \label{fig:timeorder}
    \end{subfigure}
    \begin{subfigure}[b]{0.4\textwidth}
        \includegraphics[width=\textwidth]{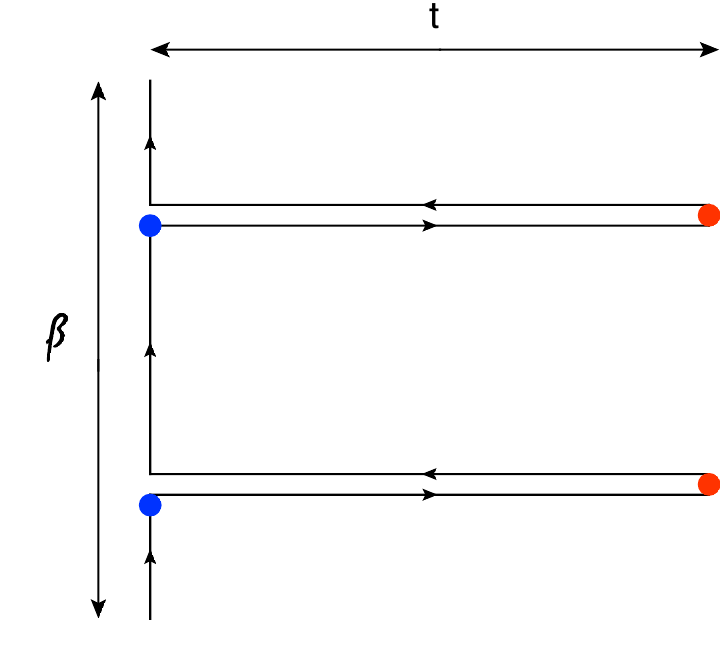}
\caption{}
        \label{fig:otoc}
    \end{subfigure}
    \caption{\ref{fig:timeorder} represents the time ordered correlation function on the two fold contour and \ref{fig:otoc} represents the OTOC. The red represent $A^{\mu}$ and blue represents $A^{\nu}$. These are two of the terms in the squared commutator for the photon field. There are two more times, one which is time ordered and one which is out of time order.}
\end{figure}

In particular, chaos exponents have been computed for several interesting field theories including scalar matrix theory, the O(N) model, the critical fermi surface, the diffusive metal, and graphene \cite{PhysRevB.98.045102}. In this paper, we study many-body chaos in $\text{QED}_{3}$ with $N_{f}$ flavors of massless fermions in the large $N_f$ limit. This theory is described by the Lagrangian
    \begin{equation}
\mathcal{L}=-\frac{1}{4e^{2}N_{f}}F^{2}+\sum^{N_{F}}_{a=1}i\bar{\psi}_{a}(x)\gamma^{\mu}D_{\mu}\psi_{a}(x)\label{eq:minkowskil}
\end{equation}
where we define the covariant derivative $D_{\mu}=\partial_{\mu}-\frac{iA_{\mu}}{\sqrt{N_{F}}}$
and the Lorentzian Dirac matrices are given by $\gamma^{\mu}=(\sigma^{z},i\sigma^{x},i\sigma^{y})$ and we work in the $+--$ signature. $\text{QED}_{3}$ appears throughout condensed matter as the proposed low energy theory for several lattice systems such as the the Kagome antiferromagnet for $N_{f}=2$. It also relevant for a variety of quantum critical points \cite{u(1)} \cite{PhysRevB.95.235146}. While these lattice models also contain monopole operators which can drastically modify the long-wavelength physics, for large enough $N_{f}$ these become irrelevant and the theory flows to a strongly coupled renormalization group fixed point. It is known that the large $N_{f}$ expansion breaks down below some critical value of $N_{f}$, below which the theory is confining \cite{Pisarski:1984dj}\cite{Appelquist:1988sr}\cite{Gusynin:2016som}\cite{Raviv:2014xna}.

Our motivations for studying this model are the following. First, we wanted to understand possible subtleties in the quantum Lyapunov exponent story associated with gauge invariance. In particular, the anti-commutator by itself is not gauge invariant, although the squared anti-commutator is. The anti-commutator can be made gauge invariant by attaching a Wilson line, so this non-locality might conceivably modify the chaos story. Second, it is interesting to understand how to detect failures of the large $N_f$ expansion. One idea is that the chaos bound of Ref.~\cite{Maldacena:2015waa} might be in tension with the large $N_f$ result for $\lambda_L$, indicating a transition. Of course, the leading $1/N_f$ term is not reliable away from large $N_f$, but it might still be suggestive of an issue. Third, we wanted to compare chaos in conventional and unconventional quantum critical points, for example, how does chaos in the O(N) model compare to chaos in $\text{QED}_{3}$? 

To analyze chaos in $\text{QED}_{3}$, we choose to study the index averaged squared anti-commutator
  \begin{equation}
\mathcal{F}(t,\x)=\frac{1}{N^{2}_{f}}\sum^{N_{F}}_{a,b=1}\text{Tr}\Big[\sqrt{\hat{\rho}}\lbrace\psi_{a}(t,\x),\bar{\psi}_{b}(0)\rbrace\sqrt{\hat{\rho}}\lbrace\psi_{a}(t,\x),\bar{\psi}_{b}(0)\rbrace^{\dagger}\Big].
\end{equation}
We restrict attention to the spatial average of $\mathcal{F}$ to avoid issues associated with spatial propagation and focus on chaos within the large local Hilbert space. However, our results could be easily extended to the compute the butterfly velocity at large $N_f$.

We briefly summarize our main results. We numerically calculate the photon polarization bubble in real time and analyze analytically it in the limit of high temperature and the limit of low frequency. We use it to numerically calculate the fermion decay rate which scales as
\begin{eqnarray}
\Gamma_{\k}\sim \frac{T}{N_{F}}
\end{eqnarray}
and vanishes as $\k\rightarrow 0$. The Lyapunov exponent takes the form
\begin{equation}
    \lambda_{L}=\frac{2\pi T}{N_{F}}\mathcal{C}
\end{equation}
with $\mathcal{C}\approx0.65$. This form is analogous to the form of $\lambda_{L}$ in the O(N) model with here the value of $\mathcal{C}$ is slightly higher. Because $\mathcal{C}<1$, the chaos bound is satisfied for all values of $N_{F}$ despite the known breakdown of the large $N_{F}$ expansion. Also, chaos is not significantly stronger in this unconventional quantum critical point as compared to the usual O(N) critical theory, at least at large $N$.

\textbf{Notation guide:}
Throughout we will use lower case letters to denote Lorentz vectors $k^{\mu}=(k^{0},\k)$ where the inner product is defined as 
\begin{equation}
  k^{2}=(k^{0})^{2}-\k^{2}  
\end{equation}
For Euclidean vectors we will use capital letter $K_{\mu}=(k_{n},\k)$ vectors and the Euclidean inner product is defined as  
\begin{equation}
 K^{2}=k^{2}_{n}+\k^{2} 
\end{equation}
\\
Integrals in Minkowski space are defined as
\begin{subequations}
\begin{align}
\int_{x}&\equiv \int \mathrm{d}t\int_{\x}
\\
\int_{k}&\equiv	\int\frac{\mathrm{d}k^{0}}{2\pi}\int_{\k}
\end{align}
\end{subequations}
Integrals and sums in Euclidean space are defined as are defined
\begin{subequations}
\begin{align}
\int_{X}&\equiv \int \mathrm{d}\tau\int_{\x}
\\
\SumInt_{K}&\equiv \frac{1}{\beta}\sum_{k_{n}}\int_{\k}
\end{align}
\end{subequations}
For Feynman slash notation $\slashed{k}\equiv \gamma^{\mu}k_{\mu}$, we will use the shorthand  $\slashed{k}^{0}=\gamma^{0}k_{0}$, $\slashed{\k}=\gamma^{i}k_{i}$ and  $\slashed{k}=\gamma^{0}k_{0}+\gamma^{i}k_{i}=\gamma^{0}k_{0}-\bm{\gamma}\cdot\k$ where $\bm{\gamma}=(i\sigma^{x},i\sigma^{y})$. In Euclidean time 
$\slashed{K}\equiv \tilde{\gamma}^{\mu}K_{\mu}$
$\slashed{K}=\tilde{\gamma}^{0}ik_{n}+\tilde{\gamma}^{i}\k_{i}$ 
To go from  to $\slashed{K}$ back to $\slashed{k}$ we have  $i\slashed{K}\rightarrow\slashed{k}$. Further conventions are specified in the appendices.

\section{Preliminaries}
    In this section, we set up the formalism required to compute $\mathcal{F}(t,\x)$ in a $1/N_{f}$ expansion
The generating function corresponding to Eqn.~\ref{eq:minkowskil} is
\begin{equation}
\mathcal{Z}=\int\mathcal{D}A \mathcal{D}\psi\mathcal{D}\bar{\psi}\,\text{exp}\left(iS_{M}\right) \label{eq:minkowskiz}
\end{equation}
In the limit $e\rightarrow \infty$ , we can neglect the Maxwell term. 

It will also be useful to consider the theory in Euclidean time. Setting $it\rightarrow \tau$ and $A_{0}\rightarrow i A_{\tau}$ we obtain the following Euclidean action
\begin{equation}
\mathcal{L}_{E}=-\frac{1}{4e^{2}N_{F}}F^{2}+\sum^{N_{F}}_{a=1}\bar{\psi}_{a}(X)\tilde{\gamma}_{\mu}D_{\mu}\psi_{a}(X)\label{eq:euclideanl}
\end{equation}
where the Euclidean and Minkowski Dirac matrices are related by $\tilde{\gamma}^{0}\equiv \gamma^{0}$ and $\tilde{\gamma}^{i}\equiv -i\gamma^{i}$ and $\tilde{\gamma}^{i}=\tilde{\gamma}_{i}$. We will use repeated lower indices and capital letters when using the Euclidean metric
  This gives us the generating functional
\begin{equation}
\mathcal{Z}_{E}=\int\mathcal{D}A\mathcal{D}\psi\mathcal{D}\bar{\psi}\,\text{exp}\left(-\int_{X}\mathcal{L}_{E}\right)\label{eq:euclideanz}
\end{equation}
where $\int_{X}\equiv\int^{\beta}_{0}\mathrm{d}\tau\int_{\x}$.
\\
To compute $F(t,\x)$, we will need the Wightman, Euclidean, and Euclidean propagators for the fermions and the gauge field. For fermions these are defined as
\begin{subequations}
	\begin{align}
	\tilde{G}^{ab}(t,\x)&=\text{Tr}\left[\sqrt{\hat{\rho}} \psi_{a}(t,\x)\sqrt{\hat{\rho}}\bar{\psi}_{b}(0,\0)\right]
	\\
	G^{ab}_{E}(\tau,\x)&=\text{Tr}\left[\hat{\rho} \psi_{a}(\tau,\x)\bar{\psi}_{b}(0,\0)
	\right]
	\\
	G^{ab}_{R}(t,\x)&=-i\text{Tr}\left[\hat{\rho}\lbrace \psi_{a}(t,\x),\bar{\psi}_{b}(0,\0)\rbrace\theta(t)\right]
	\end{align}
\end{subequations}
In momentum space these propagators can be written in the spectral representation
\begin{subequations}
	\begin{align}
	G_{W}(\slashed{k})&=\frac{\slashed{\rho}(k)}{2\cosh \frac{\beta k^{0}}{2}}
	\\
	G_{E}(\slashed{K})&=\int\frac{\mathrm{d}\omega}{2\pi}\frac{\slashed{\rho}(K)}{ik_{n}-\omega}
	\\
G_{R}(\slashed{k})&=\int\frac{\mathrm{d}\omega}{2\pi}\frac{\slashed{\rho}(k)}{k^{0}-\omega+i\epsilon}
\end{align}
\end{subequations}
where we have defined the spectral function as
\begin{equation}
\slashed{\rho}(k)=-2\Im[G_{R}(\slashed{k})]
\end{equation}
For free fermions the spectral function is
\begin{equation}
	\slashed{\rho}(k^{0},\k)=\frac{\pi\slashed{k}}{E_{\k}}[\delta(k^{0}-E_{\k})-\delta(k^{0}+E_{\k})]		
\end{equation}

which gives us the retarded and Wightman propagators
\begin{subequations}
\begin{align}
G_{W}(\slashed{k})&=\sum_{s}\frac{s\slashed{k}\delta(k^{0}-s E_{\k})}{2\cosh\frac{\beta E_{\k}}{2}} \label{eq:freewightman}
\\
G_{R}(\slashed{k})&=\frac{\slashed{k}}{(k^{0}+i\epsilon)^{2}-E^{2}_{\k}}
\label{eq:freeretarded}
\end{align}
\end{subequations}
\\
Similarly, for the gauge field we define the propagators as
\begin{subequations}
	\begin{align}
		\mathcal{D}^{\mu\nu}_{W}(t,\x)&=\text{Tr}\Big[\sqrt{\hat{\rho}} A^{\mu}(t,\x)\sqrt{\hat{\rho}}A^{\nu}(0,\0)\Big]
		\\
		\mathcal{D}^{\mu\nu}_{E}(\tau,\x)&=i\text{Tr}\Big[\hat{\rho} A^{\mu}(\tau,\x)A^{\nu}(0,\0)\Big]
		\\
	\mathcal{D}^{\mu\nu}_{R}(t,\x)&=-i\text{Tr}\Big[\hat{\rho}\left[ A^{\mu}(t,\x),A^{\nu}(0,\0)\right]\theta(t)\Big]
	\end{align}
\end{subequations}
In the spectral representation we can write these as
\begin{subequations}
	\begin{align}
	\mathcal{D}^{\mu\nu}_{W}(k)&=\frac{\mathcal{A}^{\mu\nu}(k)}{2\sinh \frac{\beta k^{0}}{2}}
	\\
	\mathcal{D}^{\mu\nu}_{E}(K)&=\int\frac{\mathrm{d}\omega}{2\pi}\frac{\mathcal{A}^{\mu\nu}(\omega,\k)}{ik_{n}-\omega}
	\\
		\mathcal{D}^{\mu\nu}_{R}(k)&=\int\frac{\mathrm{d}\omega}{2\pi}\frac{\mathcal{A}^{\mu\nu}(\omega,k)}{k^{0}-\omega+i\epsilon}
\\
\end{align}
\end{subequations}
where we define the spectral function as
\begin{equation}
\mathcal{A}^{\mu\nu}=-2\Im\mathcal{D}^{\mu\nu}_{R}(k)] \label{eq:photonspectral}
\end{equation}
\begin{figure}[t]
	\centering	
	\includegraphics[width=0.23\textwidth]{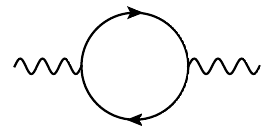}\label{fig:rungs}
	\caption{The fermion polarization bubble is the $O(1)$ contribution to the gauge field propagator and dominates over the bare term at low energies.}
\end{figure} 
The gauge field propagator $\mathcal{D}^{\mu\nu}$ is $O(1)$ in $N_{f}$ and hence a non-perturbative object. Since the self energy dominates at low energies and $\mathcal{D}^{\mu\nu}$ completely in terms of the self energy. 

At finite temperature, the choice of frame breaks Lorentz invariance. As a result, longitudinal and transverse become distinct. In $2+1d$, the photon has one longitudinal and one transverse polarization with associated polarization vectors $\epsilon^{\mu}_{L}(k)$ $\epsilon^{\mu}_{T}(k)$. A thermal mass can appear in the longitudinal component of the longitudinal part without violating gauge invariance. To make calculation simpler we will use Coulomb gauge throughout with gauge fixing condition $\nabla \cdot \vec{A}=0$. In this gauge, we define the longitudinal polarization vector $\epsilon^{\mu}_{L}(k)=u^{\mu}=(1,0,0)$, where $u^{\mu}$ is the rest frame of the medium. For $\k$ lying along the $x$ axis we can write the transverse polarization vector as following
\begin{equation}
\epsilon^{\mu}_{T}(k)=\frac{1}{\vert\k\vert}\left(0,-k^{y},
k^{x}\right)
\end{equation}
 In this gauge, we can write the bare propagator in  the following form
\begin{eqnarray}
i\mathcal{D}^{\mu\nu}(k^{0},\k)
&=&-\frac{i}{\k^{2}}u^{\mu}u^{\nu}-\frac{i}{k^{2}}P^{\mu\nu}_{T}(k) 
\label{eq:polarizationparts}
\end{eqnarray}
where $P^{\mu\nu}_{T}(k)$ is a transverse projection tensor defined as 
 \begin{equation}
P^{\mu\nu}_{T}(k)=\Big(\delta^{ij}-\frac{k^{i}k^{j}}{\vert\k\vert^{2}}\Big)g^{\mu}_{i}g^{\nu}_{j} \label{eq:transverseproj}
 \end{equation}
We can also write $P^{\mu\nu}_{T}(k)$ as a tensor product of the transverse polarization vectors
\begin{eqnarray}
P^{\mu\nu}_{T}(k)&=&\epsilon^{\mu}_{T}(\k)\epsilon^{\mu}_{T}(\k)
\end{eqnarray}
For the self energy, current conservation requires
\begin{eqnarray}
k_{\mu}\Pi^{\mu\nu}(k)&=&0
\end{eqnarray}
This allows us to write it with the following tensor structure.
\begin{eqnarray}
\Pi^{\mu\nu}(k)&=&\frac{k^{2}}{\k^{2}}\Pi^{00}(k)P^{\mu\nu}_{L}(k)+\Pi^{yy}P^{\mu\nu}_{T}(k) \label{eq:pitensor}
\end{eqnarray}
where longitudinal projector is defined as
\begin{equation}
   P^{\mu\nu}_{L}(k)=\left(-g^{\mu\nu}+\frac{k^{\mu}k^{\nu}}{k^{2}}\right)-P^{\mu\nu}_{T}(k) 
\end{equation}
Since the tensor \ref{eq:pitensor} has only two independent components, we can calculate everything in terms of the spatial components of the self energy $\Pi^{xx}(k^{0},\k)$, $\Pi^{yy}(k^{0},\k)$ and write $\Pi^{00}(k^{0},\k)$ in terms of these functions.
\\
We start with the following expression for $\mathrm{\Pi}_{\mu\nu}(K)$ in Euclidean time 
\begin{eqnarray}
\mathrm{\Pi}_{\mu\nu}(ik_{n},\k)=\SumInt_{P}\text{Tr}\left[\tilde{\gamma_{\mu}}\frac{-i\slashed{P}}{P^{2}}\tilde{\gamma}_{\nu}\frac{-i(\slashed{K}-\slashed{P})}{(K-P)^{2}}\right] \label{eq:firsexp}
\end{eqnarray}
For For Euclidean gamma matrices we have
\begin{eqnarray}
\text{Tr}[\tilde{\gamma}_{\mu}\slashed{P}\tilde{\gamma}_{\nu}\slashed{Q}]&=&- 2(P_{\mu}(K_{\nu}-P_{\nu})+P_{\nu}(K_{\mu}-P_{\mu})-\delta_{\mu\nu}(P\cdot (K-P)) \label{eq:numerator}
\end{eqnarray}
which gives us
\begin{eqnarray}
\mathrm{\Pi}_{\mu\nu}(K)&=&-2\SumInt_{\lbrace P\rbrace}\left[\frac{P_{\mu}(K-P_{\nu})+P_{\nu}(K-P_{\mu})-\delta_{\mu\nu}P\cdot(K-P)}{P^{2}(K-P)^{2}}\right] \label{eq:secondexp}
\end{eqnarray}
We can simplify the $K^{\mu}P^{\nu}+K^{\nu}P^{\mu}$ terms by shifting $P\rightarrow K-P$ in half of this term to get
\begin{eqnarray}
\mathrm{\Pi}_{\mu\nu}(K)&=&-\SumInt_{\lbrace P\rbrace}\left[\frac{2\delta_{\mu\nu}}{P^{2}}+\frac{-K^{2}\delta_{\mu\nu}+2K_{\mu}K_{\nu}-4P_{\mu}P_{\nu}}{P^{2}(K-P)^{2}}\right] \label{eq:secondexpshift}
\end{eqnarray}
We leave the rest of the calculation of the polarization bubble to appendix \ref{appendix:photonprop}.
	\begin{figure}[t]
    \centering
    \begin{subfigure}[b]{0.4\textwidth}
        \includegraphics[width=\textwidth]{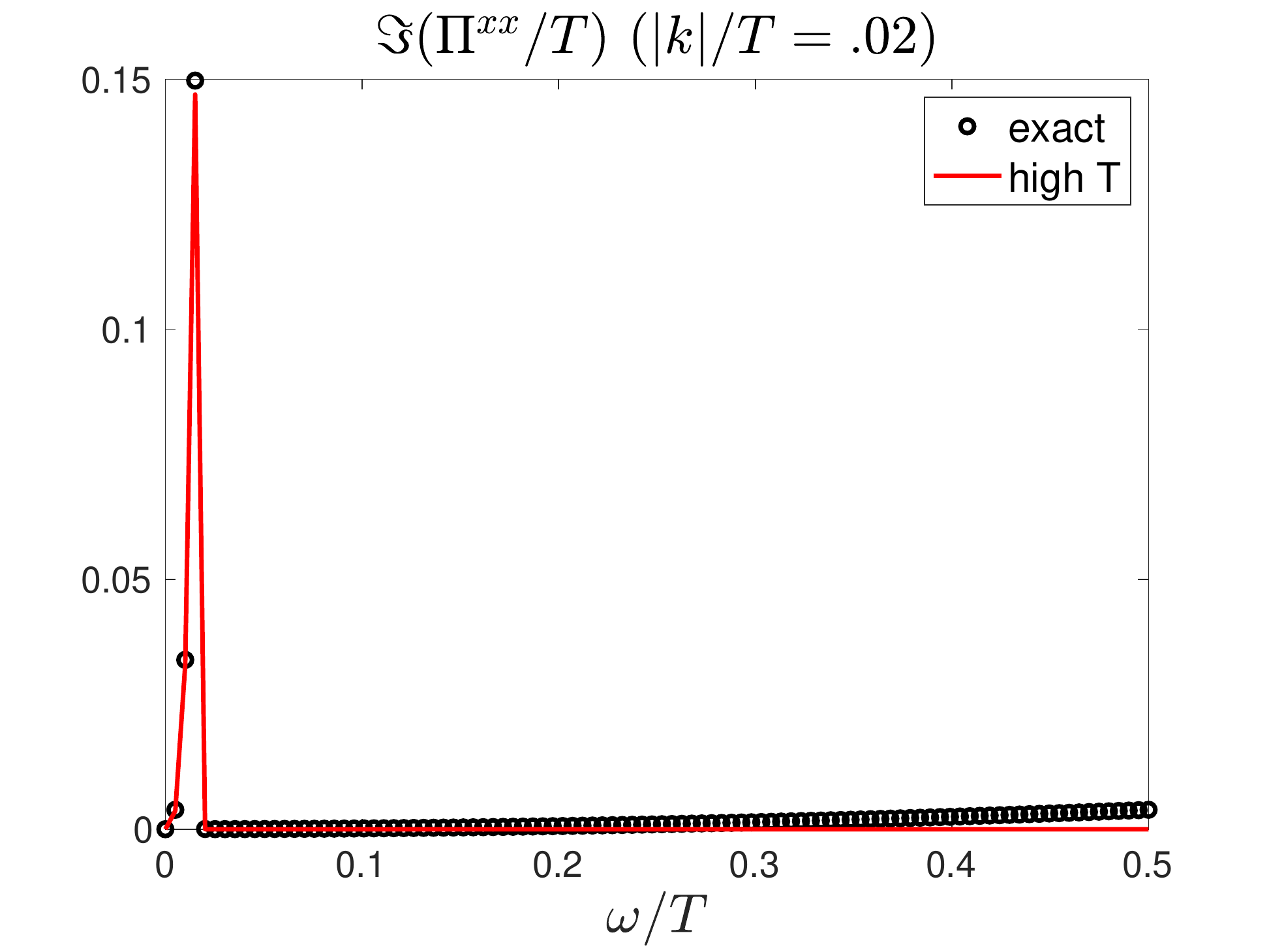}
        \caption{}
        \label{fig:impi11}
    \end{subfigure}
    \begin{subfigure}[b]{0.4\textwidth}
        \includegraphics[width=\textwidth]{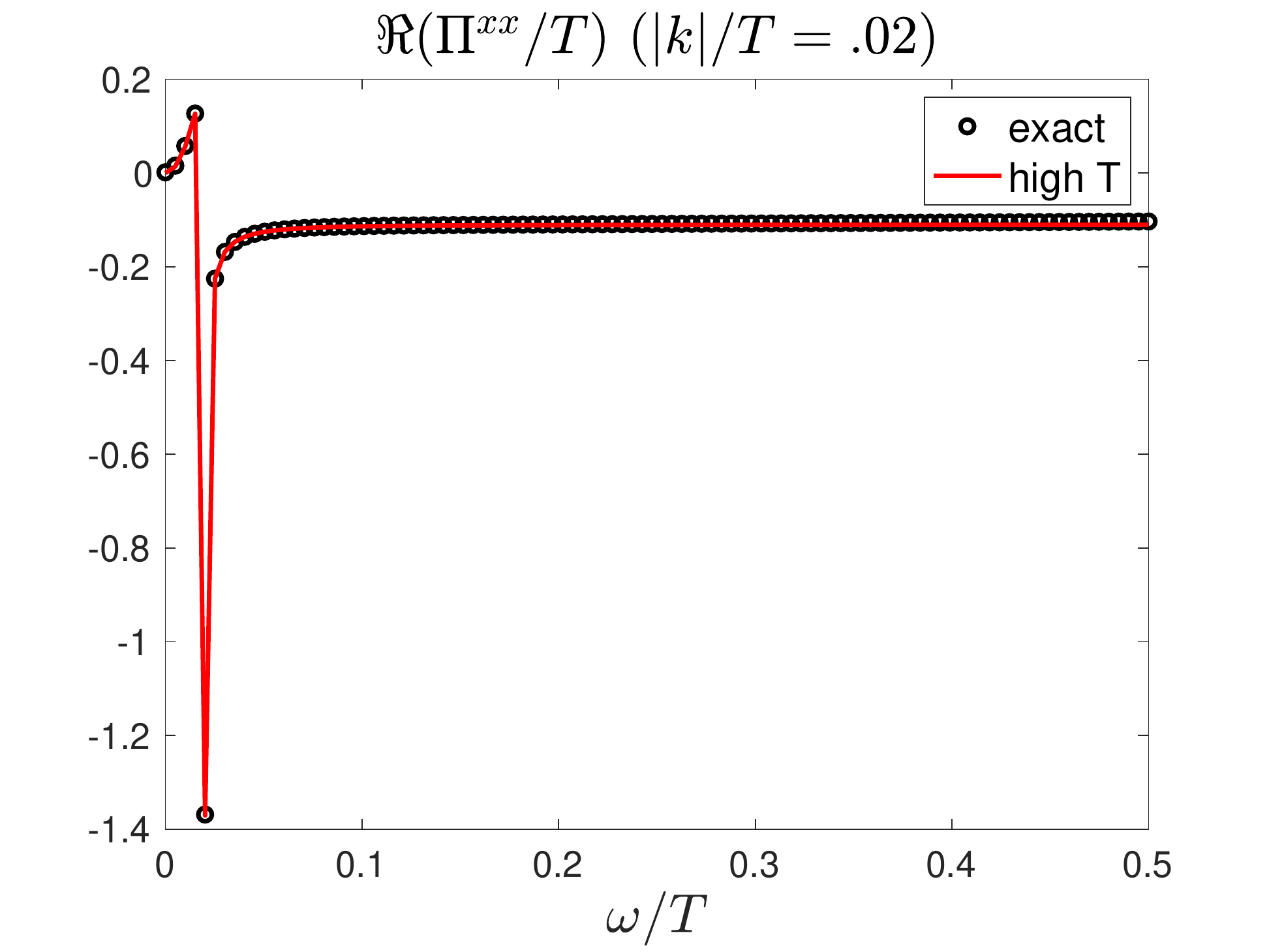}
\caption{}
        \label{fig:repi11}
    \end{subfigure}
    \caption{The imaginary and real parts of the full polarization component $\Pi^{xx}(k^{0},\k)$ are plotted against the high temperature limits in Fig.~\ref{fig:impi11} and  Fig.~\ref{fig:repi11} respectively.}
    \label{fig:pi}
    \end{figure} 
\\
We write the full photon propagator in terms of $\Pi^{00}(k^{0},\k)$ and $\Pi^{yy}(k^{0},\k)$ the as
\begin{equation}
i\mathcal{D}^{\mu\nu}(k)=-\frac{i}{\frac{1}{e^{2}}\k^{2}-\Pi^{00}(k^{0},\k)}u^{\mu}u^{\nu}-\frac{i}{\frac{1}{e^{2}}k^{2}-\Pi^{yy}(k^{0},\k)}P^{\mu\nu}_{T}(k)
\label{eq:fullphotonbare}
\end{equation}
Since the self energy is $O(1)$, it will always dominate over the bare term. Thus we drop the bare term and write 
\begin{equation}
i\mathcal{D}^{\mu\nu}(k)=\frac{i}{\Pi^{00}(k^{0},\k)}u^{\mu}u^{\nu}+\frac{i}{\Pi^{yy}(k^{0},\k)}P^{\mu\nu}_{T}(k)
\label{eq:fullphoton}
\end{equation}
At finite temperature, a thermal mass $m_{thermal}=\frac{\sqrt{e}T\log 2}{\pi}$ will appear in the longitudinal part of the propagator $\Pi^{00}$. This does not violate gauge invariance and is due to the Debye screening of the gauge field. 

Next we consider the first correction to the fermion self energy. Since the fermion mass is $1/N_{f}$ suppressed, to leading order the physical effects are controlled by the imaginary. This introduce a damping rate $\Gamma_{\textbf{k}}$ leading to a finite lifetime to the fermions. In terms of the photon and free fermion spectral functions, we can write the self energy as
\begin{eqnarray}
\Sigma(\slashed{K})&=&-
\frac{1}{N_{F}}\SumInt_{\lbrace P\rbrace}\int\frac{\mathrm{d}\nu}{2\pi}\int\frac{\mathrm{d}\omega}{2\pi} \tilde{\gamma}_{\mu}\frac{\slashed{\rho}(\omega,\p)}{ip_{n}-\omega}\tilde{\gamma}_{\nu}\frac{\mathcal{A}_{\mu\nu}(\nu,\k-\p)}{ik_{n}-ip_{n}-\nu}
\end{eqnarray}
Doing the Matsubara sum gives us
\begin{eqnarray}
\Sigma(\slashed{K})
&=&
\frac{1}{N_{F}}\int_{\p}\int\frac{\mathrm{d}\nu}{2\pi}\int\frac{\mathrm{d}\omega}{2\pi}\tilde{\gamma}_{\mu}\slashed{\rho}(\omega,\p)\tilde{\gamma}_{\nu}\mathcal{A}_{\mu\nu}(\nu,\k-\p)\Big(\frac{1-n_{F}(\omega)+n_{B}(\nu)}{ik_{n}-\omega-\nu}\Big)
\end{eqnarray}
Analytically continuing back to real time and inserting the free fermion spectral function gives us
\begin{eqnarray}
\mathrm{\Sigma}_{R}(\slashed{k})&=&\frac{1}{N_{F}}\int_{\p}\int\frac{\mathrm{d}\nu}{2\pi}\int\frac{\mathrm{d}\omega}{2\pi}\sum_{s=\pm} \frac{\pi s\delta(\omega-sE_{\p})\gamma_{\mu}\slashed{p}\gamma_{\nu}}{E_{\p}}\mathcal{A}^{\mu\nu}(\nu,\k-\p)\frac{1-n_{F}(\omega)+n_{B}(\nu)}{ik_{n}-\omega-\nu}
\nn&=&
\frac{1}{N_{F}}\int_{\textbf{p}}\frac{\mathrm{d}\nu}{2\pi}\sum_{s=\pm}\frac{s}{2E_{\p}}\gamma_{\mu}\slashed{p}_{s}\gamma_{\nu}\mathcal{A}^{\mu\nu}(\nu,\k-\p)
\frac{1-n_{F}(s E_{\p})+n_{B}(\nu)}{k^{0}-s E_{\p}-\nu+i\epsilon}
\label{eq:fermionselfenergy}
\end{eqnarray}
\begin{figure}[t]
	\centering
	\includegraphics[width=0.3\textwidth]{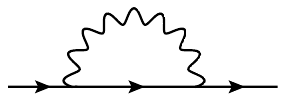}\label{fig:fermionselfenergy}
	\caption{The leading contribution to the fermion self energy is $O(\frac{1}{N_{F}})$ so the correction to the mass can be neglected.}
\end{figure}
where $p_{s}=(s E_{\p},\p)$ and $E_{\p}=\vert\p\vert$.
 Defining $\Gamma_{\k}$ as the imaginary part of the pole in the propagator, $\Gamma_{\k}$ can be written in terms of the self energy as
 \begin{equation}
 \Gamma_{\k}=-\frac{1}{2E_{\k}}\Im\text{Tr}[\slashed{k}\mathrm{\Sigma}_{R}(\slashed{k})]\Big\vert_{k^{0}=E_{\k}} \label{eq:decay}
 \end{equation}
Taking the imaginary part of Eq. \ref{eq:fermionselfenergy} we have
\begin{eqnarray}
\Im\Sigma_{R}(\slashed{k})]&=&-\frac{1}{N_{F}}\int_{\p}\int \frac{\mathrm{d}\nu}{2\pi}\sum_{s=\pm}\frac{\pi s}{2E_{\p}}\gamma_{\mu}\slashed{p}_{s}\gamma_{\nu}\mathcal{A}^{\mu\nu}(\nu,\k-\p)
\delta(k^{0}-s E_{\p}-\nu)
\nn
&&
(1-n_{F}(s E_{\p})+n_{B}(\nu))
\nn
&=&-\frac{1}{N_{F}}\int_{\p}\frac{s}{4E_{\p}}\sum_{s=\pm}\Big[\gamma_{\mu}\slashed{p}_{s}\gamma_{\nu}\mathcal{A}^{\mu\nu}(k-p_{s})
(1-n_{F}(s E_{\p})+n_{B}(k^{0}-s E_{\p}))\Big]\label{eq:fermionse}
\end{eqnarray}
Expanding the photon propagator into its longitudinal and transverse components, this becomes
\begin{eqnarray}
\label{eq:fulldecaytext}
\Gamma_{\k}&=&-\frac{1}{2E_{\k}}\Im\text{Tr}[\slashed{k}\mathrm{\Sigma}_{R}(\slashed{k})]\Big\vert_{k^{0}=E_{\k}}
\nn
&=&\frac{1}{N_{F}}\int_{\p}\sum_{s=\pm}\sum_{\lambda=L,T}\frac{s}{4E_{\k}E_{\p}}\mathcal{A}_{\lambda}(k-p_{s})
(2(k\cdot\epsilon_{\lambda}(k-p_{s}))(p_{s}\cdot\epsilon_{\lambda}(k-p_{s}))-k\cdot p_{s}(\epsilon_{\lambda}(k-p_{s})\cdot \epsilon_{\lambda}(k-p_{s})))
\nn&&
(1-n_{F}(s E_{\p})+n_{B}(k^{0}-s E_{\p}))
\Big\vert_{k^{0}=E_{\k}}\label{eq:decayrate}
\end{eqnarray}
We can evaluate Eq. \ref{eq:fulldecaytext} in Coulomb gauge by choosing the longitudinal polarization vector $\epsilon^{\mu}_{L}=u^{\mu}$. The complete expression is given in  \ref{appendix:damping}. We also note that this expression is consistent with the finite temperature optical theorem.
Implementing $\Gamma_{\k}$ as a shift in the pole, we can write the retarded propagator as
\begin{equation}
G_{R}(\slashed{k})=\frac{1}{2E_{\k}}\Big(\frac{\gamma^{0}E_{\k}-\bm{\gamma}\cdot\k}{k^{0}-E_{\k}+i\Gamma_{\k}}-\frac{-\gamma^{0}E_{\k}+\bm{\gamma}\cdot\k}{k^{0}+E_{\k}+i\Gamma_{\k}}\Big).
\end{equation}
Note that the fermion 2-point function is not gauge invariant, so this decay rate is not manifestly gauge invariant and is defined in the Coulomb gauge which we use. It is consistently used below in the $1/N_F$ expansion to compute gauge-invariant observables, and it presumably manifests in other gauge-invariant response functions.

\begin{figure}[t]
	\centering	
	\includegraphics[width=0.6\textwidth]{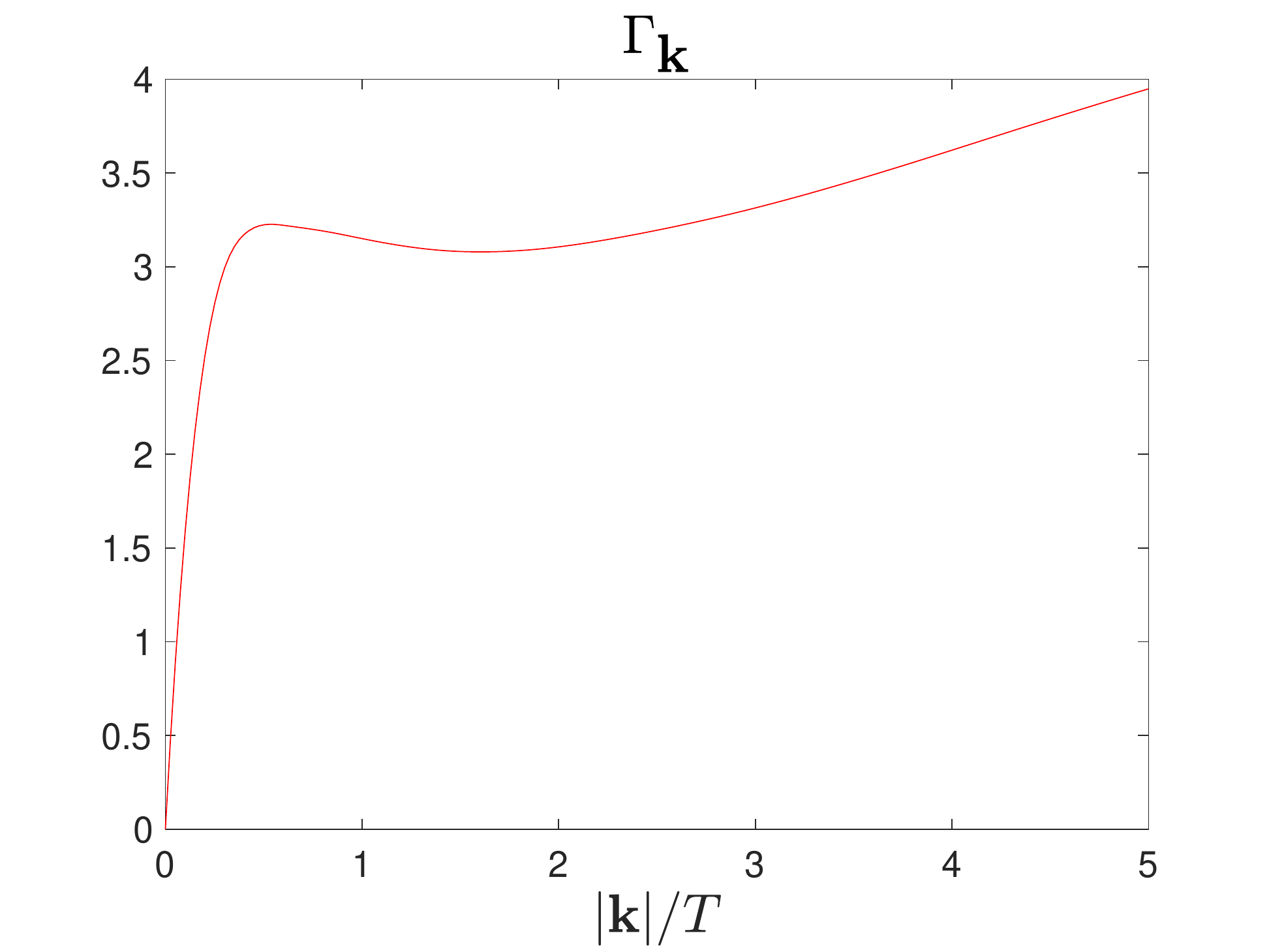}\label{fig:gamma}
	\caption{A plot of the fermion decay rate $\Gamma_{\k}$ in units of temperature which vanishes as $\k\rightarrow 0$.}
\end{figure}

\section{Diagrammatic expansion for the squared fermion anti-commutator}
We now consider the diagrammatic expansion of the squared anti-commutator of two fermion operators by expanding $\mathcal{F}(t,\x)$ in the interaction on both time folds and keeping only diagrams to order $1/N_{F}$ which are uncrossed.  The rules are very similar to those summarized in \cite{Stanford2016} except all of the propagators and hence rungs have a matrix structure. In summary:
\\
1. Vertex insertions are restricted to lie on real time folds 
\\
2. Horizontal lines correspond to retarded or advanced photon or fermions propagators including self energy corrections
\\
3. Vertical lines correspond to full photon propagators or bare fermion Wightman propagators
\\
4. Each rung or pair of rails is a matrix in spinor space and can be written in terms of a tensor product basis of gamma matrices. A rung or pair of rails is added to the ladder by matrix multiplication.
\\
We now consider the two rungs at this order. The first rung involves a single photon Wightman function sandwiched between two gamma matrices. The whole rung can be  giving us the following expression
\begin{eqnarray}
\mathcal{R}_{1}(k-k^{\prime })&=&\frac{1}{N_{F}}\gamma_{\mu}\mathcal{D}^{\mu\nu}_{W}(k-k^{\prime})\gamma_{\nu}
\nn
&=&\frac{1}{N_{F}}\gamma_{\mu}\mathcal{Q}_{B}(k^{0}-k^{\prime 0})\mathcal{A}^{\mu\nu}(k-k^{\prime})\gamma_{\nu}
\end{eqnarray}
where we define $\mathcal{Q}_{B}(\omega)=(2\sinh(\frac{\beta\omega}{2}))^{-1}$ for bosons and $\mathcal{Q}_{F}(\omega)=(2\cosh(\frac{\beta\omega}{2}))^{-1}$ for fermions.
We can write the longitudinal and transverse contributions to  $\mathcal{R}_{1}(k-k^{\prime })$ in terms of a tensor products of polarization vectors. In $2+1d$, we can express the transverse projector $P^{\mu\nu}_{T}(k)$ as the tensor product of a single transverse polarization vector
\begin{equation}
       P^{\mu\nu}_{T}(k)=\epsilon^{\mu}_{T}(\k)\epsilon^{\nu}_{T}(\k) \label{eq:polarizationproj}
    \end{equation}
    where we define $\epsilon^{\nu}_{T}(\k)$ lying along the $x$ axis as
    \begin{equation}
\epsilon^{\mu}_{T}(\k-\k^{\prime})=\frac{1}{\vert\k\vert}\left(0,- k^{y},
k^{x}\right) 
\end{equation}
Defining $\bar{k}=k-k^{\prime}$ as the relative momentum, we can write a general expression for $\mathcal{R}_{1}(\bar{k})$
\begin{eqnarray}
\mathcal{R}_{1}(\bar{k})&=&\frac{1}{N_{F}}\sum_{\lambda=L,T}
\mathcal{Q}_{B}(\bar{k}^{0})\mathcal{A}_{\lambda}(\bar{k})(\slashed{\epsilon}_{\lambda}(\bar{k})\slashed{\epsilon}_{\lambda}(\bar{k}))
\end{eqnarray}
where we have separated the longitudinal and transverse parts of the rung. Using the definitions of the longitudinal and transverse polarization vectors this becomes
\begin{eqnarray}
\mathcal{R}_{1}(\bar{k})&=&\frac{\mathcal{Q}_{B}(\bar{k}^{0})}{N_{F}}\left(\mathcal{A}_{L}(\bar{k})(\gamma_{0}\gamma_{0})+\mathcal{A}_{T}(\bar{k})(\gamma_{i}\gamma_{j})(\delta^{ij}-\hat{\bar{\k}}^{i}\hat{\bar{\k}}^{j})\right)
\end{eqnarray}

The second rung is a box structure consisting of two fermion Wightman functions and two retarded photon propagators. 
We can write this as
\begin{eqnarray}
\mathcal{R}_{2}(k,k^{\prime})&=&\frac{1}{N_{F}}\int_{p}\gamma_{\rho}G_{W}(\slashed{p}-\slashed{k}^{\prime})
\gamma_{\sigma}\gamma_{\nu}G_{W}(\slashed{k}-\slashed{p})\gamma_{\mu}\mathcal{D}^{\sigma\nu}_{R}(p)\mathcal{D}^{\mu\rho}_{R}(q-p)
\end{eqnarray}
where $p^{\mu}=(p^{0},\textbf{p})$
Inserting the free fermion spectral functions, we get
\begin{eqnarray} 
\mathcal{R}_{2}(k,k^{\prime})&=&\frac{1}{N_{F}}\int_{p}\sum_{ss^{\prime}=\pm}
\frac{\pi^{2}ss^{\prime}\delta((p^{0}-k^{\prime 0})-s^{\prime}E_{\k^{\prime}-\p})\delta((k^{0}-p^{0})-s E_{\k-\p})}{E_{\k^{\prime}-\p}E_{\k-\p}}
\nn
&&
\mathcal{Q}_{F}(p^{0}-k^{\prime 0})\mathcal{Q}_{F}(k^{0}-p^{0})
(\gamma_{\rho}(\slashed{p}-\slashed{k}^{\prime})\gamma_{\sigma}\gamma_{\nu}(\slashed{k}-\slashed{p})\gamma_{\mu})
\mathcal{D}^{\mu\rho}_{R}(q-p)\mathcal{D}^{\sigma\nu}_{R}(p)
\label{eq:firstr2expression}
\end{eqnarray}
defining the absolute Lorentz momentum $K=\frac{k+k^{\prime}}{2}$ corresponding to the Lorentz vector $K^{\mu}=(K^{0},\K)$ and the relative momentum $\bar{k}=k-k^{\prime}$ corresponding to the Lorentz vector $\bar{k}^{\mu}=(\bar{k}^{0},\bar{\k})$. Shifting $p\rightarrow p+K$, we define $E_{\pm}=E_{\bar{\k}/2\pm\p}$ given by
\begin{eqnarray}
E_{\pm}=\sqrt{\vert\bar{\k}\vert^{2}/4+\vert\p\vert^{2}\pm \vert\bar{\k}\vert\vert\p\vert\cos\theta}
\end{eqnarray} 
where $\theta$ is the angle between $\bar{\k}$ and $\p$
This gives us
\begin{eqnarray} 
\mathcal{R}_{2}(k,k^{\prime})&=&\frac{1}{N_{F}}\int_{p}\sum_{ss^{\prime}=\pm}
\frac{\pi^{2}ss^{\prime}\delta(\bar{k}^{0}/2+p^{0}-sE_{+})\delta(\bar{k}^{0}/2-p^{0}-s^{\prime} E_{-})}{E_{+}E_{-}}
\nn
&&
\mathcal{Q}_{F}(-\bar{k}^{0}/2-p^{0})\mathcal{Q}_{F}(\bar{k}^{0}/2-p^{0})
(\gamma_{\rho}(\bar{\slashed{k}}/2+\slashed{p})\gamma_{\sigma}\gamma_{\nu}(\bar{\slashed{k}}/2-\slashed{p})\gamma_{\mu})
\mathcal{D}^{\mu\rho}_{R}(q-p-K)\mathcal{D}^{\sigma\nu}_{R}(p+K)
\end{eqnarray}
performing the $p^{0}$ integral in Eq. \ref{eq:firstr2expression}, we get
\begin{eqnarray} 
	\mathcal{R}_{2}(\bar{k},K)
&=&
 \frac{1}{N_{F}}\int_{\p}\sum_{ss^{\prime}=\pm}\frac{\pi ss^{\prime}\delta(\bar{k}^{0}-sE_{+}-s^{\prime}E_{-})}{2E_{+}E_{-}}
\mathcal{Q}_{F}(s^{\prime} E_{-})\mathcal{Q}_{F}(\bar{k}^{0}- s^{\prime} E_{-})
\nn 
&&
(\gamma_{\rho}(\bar{\slashed{k}}/2+\tilde{\slashed{p}})\gamma_{\sigma}\gamma_{\nu}(\bar{\slashed{k}}/2-\tilde{\slashed{p}})\gamma_{\mu})
\mathcal{D}^{\sigma\nu}_{R}(\tilde{p}+K)\mathcal{D}^{\mu\rho}_{R}(-\tilde{p}-K)
\label{eq:finalgammarung}
\end{eqnarray}
where $\tilde{p}^{\mu}=(\bar{k}^{0}/2-s^{\prime} E_{-},\p)$. In three dimensions $\gamma^{\mu}\gamma^{\nu}\gamma^{\rho}=g^{\mu\nu}\gamma^{\rho}-g^{\mu\rho}\gamma^{\nu}+g^{\nu\rho}\gamma^{\mu}$ which allows us to write Eqn.~\ref{eq:finalgammarung} in terms of the polarization vectors as follows
\begin{eqnarray} 
\mathcal{R}_{2,\lambda\lambda^{\prime}}(\bar{k},K)&=&\frac{1}{N_{F}}\int_{\p}\sum_{\lambda\lambda^{\prime}=L,T}\sum_{ss^{\prime}=\pm}\frac{\pi ss^{\prime}\delta(\bar{k}^{0}-s E_{+}-s^{\prime}E_{-})}{2E_{+}E_{-}}
\mathcal{Q}_{F}(s^{\prime} E_{-})\mathcal{Q}_{F}(\bar{k}^{0}- s^{\prime} E_{-})
\nn
&&
((\bar{k}+2\tilde{p})\cdot\epsilon_{\lambda^{\prime}}(-\tilde{p}-K)\slashed{\epsilon}_{\lambda}(\tilde{p}+K)-(\bar{\slashed{k}}/2+\tilde{\slashed{p}})\epsilon_{\lambda}(\tilde{p}+K)\cdot\epsilon_{\lambda^{\prime}}(-\tilde{p}-K)
\nn
&&((\bar{k}-2\tilde{p})\cdot\epsilon_{\lambda}(\tilde{p}+K)\slashed{\epsilon}_{\lambda^{\prime}}(-\tilde{p}-K)
-(\bar{\slashed{k}}/2-\tilde{\slashed{p}})\epsilon_{\lambda}(\tilde{p}+K)\cdot\epsilon_{\lambda^{\prime}}(-\tilde{p}-K))
\nn 
&&\mathcal{D}_{R\lambda}(\tilde{p}+K)\mathcal{D}_{R\lambda^{\prime}}(-\tilde{p}-K)
\end{eqnarray}
Since the external momentum $\q$ is $O\left(\frac{1}{N_{F}}\right)$, the momenta of the two photon propagators are approximately anti-parallel. Therefore, we only need to consider terms in the sum over $\lambda$ where $\lambda=\lambda^{\prime}$.
\\
We can now write the whole rung in the tensor product form
\begin{eqnarray}
\mathcal{R}_{2}(\bar{k},K)
&=&\frac{1}{N_{F}}\int_{\p}\sum_{ss^{\prime}=\pm}\sum_{\lambda=L,T}\frac{\pi ss^{\prime}\delta(\bar{k}^{0}-s E_{+}-s^{\prime}E_{-})
}{2E_{+}E_{-}}
\mathcal{Q}_{F}(s^{\prime} E_{-})\mathcal{Q}_{F}(\bar{k}^{0}- s^{\prime}
E_{-})
\nn
&&\mathcal{D}_{R,\lambda}(\tilde{p}+K)\mathcal{D}_{R,\lambda}(-\tilde{p}-K)\left(\mathcal{R}^{(1)}_{2,\lambda}(\bar{k},K,\p)^{\mu}\gamma_{\mu}\mathcal{R}^{(2)}_{2,\lambda}(\bar{k},K,\p)^{\nu}\gamma_{\nu}\right)
\label{eq:generalrung2}
\end{eqnarray}
where we define the coefficients of $\gamma_{\mu}\otimes\gamma_{\nu}$ as
\begin{eqnarray}
\mathcal{R}^{\mu\nu}_{2}(\bar{k},K)
&=&\frac{1}{N_{F}}\int_{\p}\sum_{\pi ss^{\prime}=\pm}\sum_{\lambda=L,T}\frac{ss^{\prime}\delta(\bar{k}^{0}-s E_{+}-s^{\prime}E_{-})
}{2E_{+}E_{-}}
\mathcal{Q}_{F}(s^{\prime} E_{-})\mathcal{Q}_{F}(\bar{k}^{0}- 
E_{-})
\nn
&&\mathcal{D}_{R,\lambda}(\tilde{p}+K)\mathcal{D}_{R,\lambda}(-\tilde{p}-K)\mathcal{R}^{(1)}_{2,\lambda}(\bar{k},K,\p)^{\mu}\mathcal{R}^{(2)}_{2,\lambda}(\bar{k},K,\p)^{\nu}
\label{eq:numeratorcoefficients}
\end{eqnarray}
The expressions for the matrix elements defined in \ref{eq:numeratorcoefficients} are contained in appendix.
\\
\\
We can do the integral over $\vert \p\vert$ where the zeroes of the delta function are given by
\begin{equation}
\vert\p\vert_{0}=\frac{\bar{k}^{0}}{2}\sqrt{\frac{(\bar{k}^{0})^{2}-\vert\bar{\k}\vert^{2}}{(\bar{k}^{0})^{2}-\vert\bar{\k}\vert^{2}\cos^{2}\theta_{\p}}}
\end{equation}
where $\theta_{\p}$ is the angle between the loop momenta $\p$ and the relative momenta $\bar{\k}$. 
\\
For the root to be positive, we have condition $(k^{0})^{2}<\vert\k\vert^{2}\cos^{2}\theta$ or $(k^{0})^{2}>\vert\k\vert^{2}$, which occurs for $(s,s^{\prime})=(1,1)$ and $(s,s^{\prime})=(\pm 1, \mp 1)$.
\section{Bethe Salpeter equation and the Lyapunov exponent}
Our diagrammatic expansion of the squared fermion anti-commutator will give us an integral equation we will solve to obtain the Lyapunov exponent and butterfly velocity. This equation is reminiscent of the original Bethe-Salpeter equations first used to study bound states in quantum electrodynamics in \cite{gellmannlow}\cite{bethesalpeter}\cite{salpeter}. 
\\
In matrix form, the equation for the ladder is basis dependent. However 
it, turns out that at long times, we can take advantage of the matrix structure of this equation and turn it into two scalar equations for components of a ``wave-function"
To illustrate this, we start define the Lorentz vectors $q^{\mu}=(\nu,\q)$ $k^{\mu}=(k^{0},\k)$ and the following on-shell projectors 
\begin{equation}
\Lambda^{(i)}_{\pm}(\k)=\frac{E_{\k}\pm\gamma^{0}\bm{\gamma}\cdot\k}{2E_{\k}} \label{eq:proj}
\end{equation}
where $i=1,2$ refers to the particle index and $\bm{\gamma}\cdot\k=-\gamma^{i}k_{i}$. 
These have the following properties
\begin{equation}
    \Lambda^{(i)}_{\pm}(\k)\Lambda^{(i)}_{\pm}(\k)=\Lambda^{(i)}_{\pm}(\k),\enspace\enspace  \Lambda^{(i)}_{\pm}(\k)\Lambda^{(i)}_{\mp}(\k)=0, \enspace\enspace \Lambda^{(i)}_{+}(\k)+\Lambda^{(i)}_{-}(\k)=\mathbb{1}\label{eq:projid} 
\end{equation}
The eigenvectors of $\Lambda^{(i)}_{\pm}(\k)$ are
\begin{subequations}
	\begin{align}
	u_{+}(k) &=\left(\frac{E_{\k}}{ik^{x}+k^{y}},1
\right) \label{eq:onshelleigvec1}
	\\
	u_{-}(k) &= \left(-\frac{E_{\k}}{ik^{x}+k^{y}},1
\right) \label{eq:onshelleigvec3}
	\end{align}
	\label{eq:onshelleigvec1,eq:onshelleigvec3}
\end{subequations}
From Eqns. \ref{eq:onshelleigvec1} and \ref{eq:onshelleigvec3} we see $u_{-}(\k)=u_{+}(-\k)$.
Acting with the projectors on $u_{\pm}(\k)$ we get
\begin{subequations}
\begin{align}
\Lambda_{\pm}(\k)u_{\pm}(\k)=u_{\pm}(\k)\enspace\enspace\Lambda_{\mp}(\k)u_{\pm}(\k)=0
\end{align}
\end{subequations}
From this, we conclude that $u_{\pm}(\k)$ are eigenvectors of $\Lambda_{\pm}(\k)$ with eigenvalue one and $u_{\pm}(\k)$ are an eigenvectors of $\Lambda_{\mp}(\k)$ with eigenvalue zero.
We define $\Lambda^{(1)}_{s}(\k)\Lambda^{(2)}_{s^{\prime}}(\k)$ as the tensor product of two projection operators. Eq. \ref{eq:projid} gives us the additional four properties
\begin{equation}
(\Lambda^{(1)}_{+}(\k)\Lambda^{(2)}_{+}(\k)+(\Lambda^{(1)}_{-}(\k)\Lambda^{(2)}_{+}(\k)+(\Lambda^{(1)}_{+}(\k)\Lambda^{(2)}_{-}(\k))+(\Lambda^{(1)}_{-}(\k)\Lambda^{(2)}_{-}(\k))=\mathbb{1} \label{eq:tensorprojectorid}
\end{equation}
The eigenvectors of $u^{(1,2)}_{\pm}$. $\Lambda^{(1)}_{s}(\k)\Lambda^{(2)}_{s^{\prime}}(\k)$ can be written as tensor products of the form
\begin{equation}
u^{(1,2)}_{ss^{\prime}}(\k)=u^{(1)}_{s}(\k)u^{(2)}_{s^{\prime}}(\k)
\end{equation}
We can use the free fermion spectral function to write the retarded Green's
function in the form $\slashed{k}g_{R}(k)$. 
\begin{equation}
G^{\ast}_{R}(\slashed{k}-\slashed{q}) G_{R}(\slashed{k})=(\slashed{k}-\slashed{q}) \slashed{k}g^{\ast}_{R}(k-q)g_{R}(k)
\end{equation}
Consider the product $g^{\ast}_{R}(q-k)g_{R}(k)$. Writing it explicitly in terms of momentum and frequency arguments we have
\begin{eqnarray}
g^{\ast}_{R}(k-q)g_{R}(k)&=&
\sum_{ss^{\prime}=\pm}\frac{ss^{\prime}}{4E_{\k}E_{\k-\q}}\left(\frac{1}{k^{0}-\nu-sE_{\k-\q}-i\epsilon}\right)\left(\frac{1}{k^{0}-s^{\prime}E_{\k}+i\epsilon}\right) \label{eq:grtimesgr}
\end{eqnarray}
Evaluating the $k^{0}$ integral we have four poles: $k^{0}=\nu-sE_{\k-\q}+i\epsilon$ and $k^{0}=s^{\prime}E_{\k}-i\epsilon$
which give residues proportional to $\sim(\nu+sE_{\k-\q}-s^{\prime} E_{\k})^{-1}$. To get the longtime behavior, we keep only the contribution from the most singular terms $\sim(\nu\pm E_{\k-\q}\mp E_{\k})^{-1}$ where we close the contour clockwise on the lower half plane
\\
In the end we have
\begin{eqnarray}
g^{\ast}_{R}(k-q)g_{R}(k)&\rightarrow&
\sum_{s=\pm}-\frac{\pi i\delta(k^{0}-sE_{\k})}{2E_{\k-\q}E_{\k}}\left[\frac{1}{\nu-s(E_{\k}-E_{\k-\q})+i\epsilon}\right] \label{eq:grprod}
\end{eqnarray}
Note that for $\q=0$, these terms become proportional to $\nu^{-1}$.
Assuming $\Gamma_{\q-\k}\approx\Gamma_{\q}$ we can shift the poles in \ref{eq:grprod} to get
\begin{eqnarray}
g^{\ast}_{R}(k-q)g_{R}(k)&\rightarrow&
\sum_{s=\pm}\frac{\pi i\delta(k^{0}-sE_{\k})}{2E_{\k-\q}E_{\k}}\left[\frac{1}{\nu-s(E_{\k}-E_{\k-\q})+2i\Gamma_{\k}}\right]
\end{eqnarray}
For small $\q$, we can express Eqn.~\ref{eq:grprod} in terms of projection operators as follows
\begin{eqnarray}
G^{\ast}_{R}(\slashed{k}-\slashed{q})G_{R}(\slashed{k})&\approx&\sum_{s=\pm}s2\pi i\delta(k^{0}-sE_{\k})\frac{\Lambda^{(1)}_{s}(\k)\gamma^{0}\Lambda^{(2)}_{s}(\k)\gamma^{0}}{\nu-s(E_{\k}-E_{\k-\q})+2i\Gamma_{\k}}\label{eq:grproj}
\end{eqnarray}
where we used using $(\gamma^{0})^{2}=\mathbb{1}$.
Consider the first term in the ladder, which is the tensor product of two retarded Green's functions
\begin{equation}
\mathcal{F}_{0}(t,\x)=\frac{1}{N_{F}}[G_{R}(t,\x)G^{\ast}_{R}(t,\x)]
\end{equation}
In momentum space we can define the object $f(q,k)$ as
\begin{equation}
\mathcal{F}(q)=\frac{1}{N_{F}}\int_{k}f(q,k)
\end{equation}
And write the first term as
\begin{equation}
\mathcal{F}_{0}(q)=\frac{1}{N_{F}}\int_{k}G^{\ast}_{R}(\slashed{k}-\slashed{q}) G_{R}(\slashed{k}) \label{eq:firstterm}
\end{equation}
We can now write the infinite ladder as
\begin{equation}
\mathcal{F}(q)=\frac{1}{N_{F}}\int_{k} G^{\ast}_{R}(\slashed{k}-\slashed{q}) G_{R}(\slashed{k})\left(\mathbb{1}+\int_{k^{\prime}}\mathcal{R}(k,k^{\prime})f(k,k^{\prime},q)\right)
\label{eq:secondterm}
\end{equation}
Where $q^{\mu}=(\nu,\q)$, $k^{\mu}=(k^{0},\k)$ and $k^{\prime\mu}=(k^{\prime0},\k^{\prime})$, and $\mathcal{R}(k,k^{\prime})$ is the sum of the two rungs. It is important to note that we have computed the two rungs in a different basis, and we would need to use the Fierz identities to explicitly write $\mathcal{R}(k,k^{\prime})$  in the correct matrix form.
\\
Assuming exponential growth at long times, the ladder will remain invariant under the addition of one rung. For $\q$ small, we can make the approximation $E_{\k}-E_{\k-\q}= \delta E_{\q}\approx \q \cdot v_{\k}$ where $v_{\k}=\nabla_{\k}E_{\k}$ is the group velocity.
Dropping the constant term and inserting Eqn.~\ref{eq:grproj} into \ref{eq:secondterm}, we get
\begin{equation}
(-i\nu+i\delta E_{\q}+2\Gamma_{\k})f(q,k)=\frac{1}{N_{F}}\int_{k^{\prime}}\sum_{s=\pm}(\Lambda^{(1)}_{s}(\k)\gamma^{0}\Lambda^{(2)}_{s}(\k)\gamma^{0})\delta(k^{0}-E_{\k}) \mathcal{R}(k,k^{\prime})f(q,k^{\prime})
\label{eq:firsttermexponential}
\end{equation}
Acting on both sides of Eqn.~\ref{eq:firsttermexponential} with $(\Lambda^{(1)}_{s}(\k)\gamma^{0}\Lambda^{(2)}_{s}(\k)\gamma^{0})$, we see from Eqn.~\ref{eq:tensorprojectorid} that Eqn.~\ref{eq:firsttermexponential} is just multiplied by one or zero. So instead of considering the matrix $f(q,k)$, we can consider an eigenvalue equation for a vector $v(q,k)$ which is an eigenvector of the projection operators. 
We can now propose the following on-shell form of $v(q,k)$
\begin{eqnarray}
v(\nu,\q,\k)=\sum_{s=\pm}\Lambda^{(1)}_{s}(\k)\Lambda^{(2)}_{s}(\k)\delta(k^{0}-s E_{\k})v_{s}(\nu,\q,\k)\label{eq:onshellpsi}
\end{eqnarray}
\begin{figure}[t]
	\centering
	\includegraphics[width=0.9\textwidth]{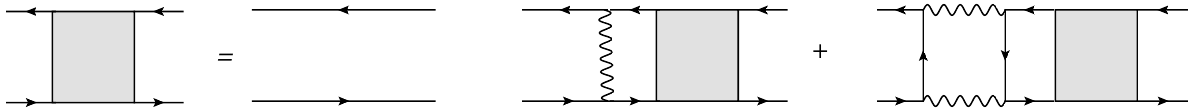}\label{fig:fermionanticommutator}
	\caption{A diagrammatic picture of the square of the fermion anticommutator. The lower rails and retarded propagators and upper rails are advanced propators and the rails are Wightman functions.}
\end{figure}
Since $v_{\pm}(\nu,\q,\k)$ are eigenstates of $\Lambda^{(1)}_{\pm}(\k)\Lambda^{(2)}_{\pm}(\k)$ for small $q$ we can write $v^{(i)}_{\pm}(\nu,\q,\k)$ in the form
 \begin{equation}
 v_{\pm}(\nu,\q,\k) = u^{(1,2)}_{\pm}(\k)\psi_{\pm}(\nu,\q,\k) \label{eq:twoparticlewavefunction}
    \end{equation}    
where $\psi_{\pm}(\nu,\q,\k)$ is a scalar function.
To write equations for $\psi_{\pm}(\nu,\q,\k)$  we take the inner product of both sides, defining a scalar kernel written in terms of both rungs and as 
\begin{eqnarray}
K_{ss^{\prime}}(\k,\k^{\prime})=\frac{
\langle u^{(1,2)}_{s}(\k)|\gamma^{0}_{(1)}\gamma^{0}_{(2)}(\mathcal{R}_{1}(s\k,s^{\prime}\k^{\prime})+\mathcal{R}_{2}(s\k,s^{\prime}\k^{\prime}))|u^{(1,2)}_{s^{\prime}}(\k^{\prime})\rangle}{\langle u^{(1,2)}_{s}(\k)\vert u^{(1,2)}_{s}(\k)\rangle}
\label{eq:scalarkernel}
\end{eqnarray}
Because inner product is basis independent, it does not matter that the rungs are written in different bases.
\begin{eqnarray}
(-i\nu+is\delta E_{\q}+2\Gamma_{\k})\psi_{s}(\nu,\q,\k)&=&\sum_{s^{\prime}=\pm}\frac{1}{N_{F}}
\int_{\k^{\prime}}K_{ss^{\prime}}(\k,\k^{\prime})\psi_{s^{\prime}}(\nu,\q,\k^{\prime})  \label{eq:bethesalpeter}
\end{eqnarray}
Eqns. \ref{eq:bethesalpeter} now looks like two coupled effective Dirac equations for a scalar wave function.
We can now define a kernel matrix in the $ss^{\prime}$ indices as 
\begin{eqnarray}
\hat{K}_{ss}&=&K_{ss}-2N_{F}\Gamma_{\k}\delta_{ss^{\prime}}
\end{eqnarray}
and $\Psi(\nu,\q,\k)=(\psi_{+}(\nu,\q,\k),\psi_{-}(\nu,\q,\k)^{T}$
\begin{eqnarray}
(-i\nu\hat{\sigma}_{0}+i\delta E_{\q}\hat{\sigma}_{z})\Psi(\nu,\q,\k)&=&\frac{1}{N_{F}}
\int_{\k^{\prime}}\hat{K}(\k,\k^{\prime})\Psi(\nu,\q,\k^{\prime})  \label{eq:bethesalpeter2}
\end{eqnarray}
In the time domain, this becomes
\begin{eqnarray}
(\partial_{t}\hat{\sigma}_{0}+i\q\cdot v_{\k}\hat{\sigma}_{z})\Psi(t,\q,\k)&=&\frac{1}{N_{F}}
\int_{\k^{\prime}}\hat{K}(\k,\k^{\prime})\Psi(t,\q,\k^{\prime})  \label{eq:bethesalpetertime}
\end{eqnarray}
\section{Numerics}
We define the Lyapunov exponent to be the largest eigenvalue of $\hat{K}$ at $\q=0$ with eigenvector $\psi_{1}$. We assume $\psi_{1}$ is rotationally symmetric so the eigenvalue equation can be simplified by projecting both sides onto states with angular momentum $l = 0$. The resulting eigenvalue equation involves a one dimensional integral transform, involving only the magnitude of the momentum. Below we summarize the computational procedure, with further details in Appendix~\ref{appendix:numerics}.

First, we compute $\Pi^{xx}$ and $\Pi^{yy}$ numerically on a fine grid in $(\omega, |\k|)$ space and use these to construct the full photon spectral function matrix. We then use these to calculate the fermion decay rate and the matrix components of the rung functions. Since these quantitites effectively vanish when the norms of their arguments greatly exceed T, we use a simple linearly spaced grid with a hard momentum cutoff of order $10$ T . The temperature is the only scale in the problem, so the entire numerical computation is set up in terms of the scaled variables $\mathfrak{k}  = k/T$ and $\mathfrak{w} = \omega/T$ .

We can write $\lambda_{L}$ in the form
\begin{equation}
    \lambda_{L}=\frac{2\pi T}{N_{F}}\mathcal{C}
\end{equation}
Numerics give $\mathcal{C}\approx0.65$. $\lambda_{L}$ takes the same form for the O(N) model with $\mathcal{C}=0.51$. 
For nonzero $\q$ we can also calculate a butterfly velocity $v_{B}$ which we expect to be $v_{B}\sim 1$. We do not give a precise number as computing $v_{B}$ is more numerically involved and the spatial propagation is expected to be significantly modified by higher order in $\frac{1}{N_{F}}$ corrections
\cite{Xu2018a}.

\section{Discussion}

	\begin{figure}[t]
    \centering
    \begin{subfigure}[b]{0.3\textwidth}
        \includegraphics[width=\textwidth]{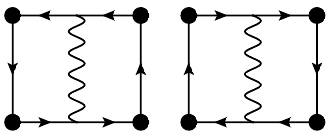}
        \caption{}
        \label{fig:current1}
    \end{subfigure}
    \begin{subfigure}[b]{0.3\textwidth}
        \includegraphics[width=\textwidth]{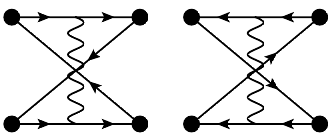}
\caption{}
        \label{fig:current2}
    \end{subfigure}
    \caption{We get four types of terms from expanding the squared current-current commutator to first order. The first nontrivial diagrams can be divided into two group and are shown in \subref{fig:current1} and \subref{fig:current2}. The black dots are current insertions. Each group contains a diagram corresponding to each direction of particle flow. Only the diagrams with uncrossed rungs will contribute to chaos.}
    \label{fig:currents}
    \end{figure} 
    
In this work we studied thermalization and chaos in $\text{QED}_{3}$ at finite temperature in a large $N_{F}$ expansion. We calculated the photon polarization tensor in real time numerically and also obtained exact expressions in the limits of high temperature and high frequency. We numerically computed the fermion decay rate and obtained the Lyapunov exponent from a two-body Dirac-like equation for the squared anti-commutator.

While the object we have computed, $\mathcal{F}(t,\x)$, is gauge invariant, the fermion anti-commutator by itself is not.
Furthermore, because of their anti-commutation relations, fermions are inherently non-local, thus the squared anti-commutator we are computing is actually a non-local object even though the growth is causal. 

We can compare the growth of the squared fermion anti-commutator with the growth of local operators. The most natural local operator in $\text{QED}_{3}$ is the fermion current. This is defined as
\begin{equation}
    j^{\mu}_{a}(t,\x)=\bar{\psi}_{a}(t,\x)\gamma^{\mu}\psi_{a}
\end{equation}
We can consider the squared commutator of two current operators
\begin{eqnarray}
\mathcal{C}(t,\x)=\frac{1}{N_{F}}\delta_{\mu\rho}\delta_{\nu\sigma}\text{Tr}\Big[\sqrt{\hat{\rho}}\left[ j^{\mu}_{a}(t,\x),j^{\nu}_{b}(0,0)\right]\sqrt{\hat{\rho}}\left[j^{\rho}_{a}(t,\x),j^{\sigma}_{b}(0,0)\right]^{\dagger}\Big]
\end{eqnarray}
Another object we could have considered is the squared commutator for the gauge field. Due to current conservation, we know that the photon and current correlation functions are related up to contact terms which vanish on shell.

We expect composite operators will grow as fast as elementary operators \cite{Chowdhury:2017jzb}. However, it is not guaranteed that these grow at the same rate. 
\begin{figure}[t]
	\centering	
	\includegraphics[width=0.5\textwidth]{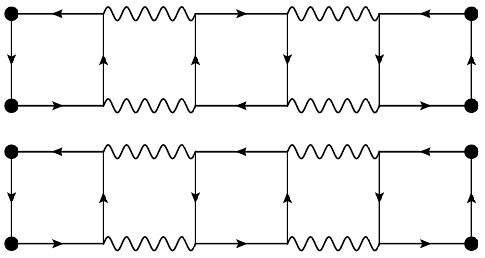}
	\caption{Terms contributing to the current current commutator at higher order.}
	\label{fig:higherorder}
	\end{figure}
We can compare $\mathcal{C}(t,\x)$ to $\mathcal{F}(t,\x)$ by looking at the structure of the ladder diagrams. The technical details are left to appendix \ref{appendix:ladders}). We see that all of the rungs in $\mathcal{F}(t,\x)$ are also contained in $\mathcal{C}(t,\x)$ as well as additional structures with crossed legs and particle arrows oriented in opposite directions.
The internal structure of this ladder contains the same terms in the squared fermion anti-commutator as well as other pieces.
If we expand $\mathcal{C}(t,\x)$ further, we obtain diagrams containing particle flows in opposing directions as shown in Fig.~\ref{fig:higherorder}. In principle, these can also occur in the ladder expansion for $\mathcal{F}(t,\x)$. However, these terms are suppressed by $\frac{1}{N_{F}}$, so they do not contribute to chaos to leading order. Thus, we argue that the diagrams in $\mathcal{F}(t,\x)$ are all contained in $\mathcal{C}(t,\x)$, so chaos must grow at least as fast as the rate obtained from $\mathcal{F}(t,\x)$.

In the future it would be interesting to see if one could adapt a kinetic method such as described in \cite{Grozdanov:2018atb} to simplify the calculation. It would also be interesting to improve numerics so that the Lyapunov exponent can be calculated to higher accuracy and get an estimate for the butterfly velocity. Our work also serves as a starting point to compute other dynamical quantities in finite temperature $\text{QED}_{3}$ such as the charge diffusion coefficient in systems with disorder. While the decay rate we computed is not gauge invariant, it is the first step in the computation of physical response functions.

\section{Acknowledgements}
We thank G. Cheng, D. Chowdhury, M. Pate, A. Patel, S. Sachdev, and S. Whitsitt for valuable disussions. We espcially thank G. Cheng for a valuable check on our calculations. JS is supported by the National Science Foundation Graduate Research Fellowship under Grant No. DGE1144152. BGS acknowledges support from the Simons Foundation as part of the It From Qubit Collaboration.

\begin{appendix}
\section{Real time Polarization bubble}
\label{appendix:photonprop}
In this appendix we will describe the calculation of the components photon polarization tensor
\\
 We start with the expression for the spatial components in Euclidean time
\beq
\Pi_{ij} =T \sum_{p_n} \int_p  \left[\frac{2 \delta_{ij}}{P^2} + \frac{-K^2 \delta_{ij} + 2 k_i k_j - 4 p_i p_j}{P^2 (K-P)^2} \right]
\eeq
where $K$ and $P$ are Euclidean vectors. Note that the numerator does not involve the summed-over Matsubara frequencies. The Matsubara sums can be performed to give
\beq
T \sum_{p_{n}} \frac{1}{p_{n}^{2} + E_{\p}^{2}} = \frac{1- 2n_F(E_{\p})}{2 E_{\p}}
\eeq
and
\beq
T \sum_{p_n} \frac{1}{p_n^2 + E_{\p}^{2}}\frac{1}{(k_{n}-p_{n})^2 + E_{k-p}^2} = \frac{1}{4 E_p E_{\k-\p}} \sum_{s,s'} \frac{s s' \left[ n_F(sE_p) + n_F(s' E_{p-k})-1 \right]}{i k_n - s E_p - s' E_{k-p}}.
\eeq
The first term, after subtracting the zero temperature value, gives
\beq
 2 \int_{p}  \frac{1- 2n_F(E_{\p})}{2 E_{\p}} - 2 \int_{p} \frac{1}{2 E_{\p}} = -\frac{T}{\pi} \int_{0}^{\infty} du \frac{1}{e^{u}+1} = -\frac{T \log 2}{\pi}.
\eeq
where $\vec{k} = (|\k|,0)$ and $\vec{p}=|\p|(\cos \theta, \sin \theta)$. The numerators of the second term are
\beq
L^{xx}= - (k_{n}^{2} + |\k|^{2})+ 2 |\k|^{2} - 4 |\p|^{2} \cos^{2} \theta
\eeq
and
\beq
L^{yy}= - (k_{n}^{2} + |\k|^{2}) - 4 |\p|^{2} \sin^{2}\theta.
\eeq

\subsection{Tensor structure}
The full real-time polarization can be expanded in two tensor structures. While the transverse component is spatially transverse, the longitudinal component is spacetime transverse. This tensor structure ensures that the full polarization bubble obeys the Ward identity required by current conservation.
\\
At finite temperature there is a new vector available set by the rest frame of the medium. The tensor structures are
\beq
P^{\mu\nu}_{L} = -g^{\mu \nu} + \frac{k^{\mu} k^{\nu}}{k^{2}}-P^{\mu \nu}_{T}
\eeq
and
\beq
P^{\mu \nu}_{T} = \delta^{\mu i} \delta^{\nu j} \left( \delta^{i j} - \frac{k^{i} k^{j}}{|\k|^{2}} \right).
\eeq
From these two structures we can define 
\beq
\tilde{P}^{\mu \nu} = P^{\mu\nu}_{L}+P^{\mu\nu}_{T}
\eeq
These obey
\beq
k_{\mu} \tilde{P}^{\mu \nu} = 0,
\eeq
\beq
k_{\mu} P^{\mu \nu}_{T} =0,
\eeq
and
\beq
k_{i} P^{ij}_{T} = 0.
\eeq
To give an invariant definition of $P^{\mu\nu}_{T}$, let $u^{\mu}$ denote the velocity of the medium, which in the medium's rest frame is $u=(1,0,0)$. Let the transverse $u$ projector be
\beq
P^{\mu \nu}_{T}(u) = - g^{\mu \nu} + \frac{u^{\mu} u^{\nu}}{u^2},
\eeq
and define
\beq
\bar{k}^{\mu} = P^{\mu \nu}_{T}(u) k_{\nu} = - k^{\mu} + (k \cdot u) u^{\mu} .
\eeq
In the medium's rest frame, $k = (k^{0}, \vec{k})$, then
\beq
\bar{k} = (0,-\vec{k}).
\eeq
The transverse projector to $\bar{k}$ is
\beq
P^{\mu \nu}_{T}(\bar{k})= - g^{\mu \nu} + \frac{\bar{k}^\mu \bar{k}^\nu}{\bar{k}^2},
\eeq
which in the medium's rest frame is
\beq
P^{00}_{L}(\bar{k}) = -1 = - u^0 u^0,
\eeq
\beq
P^{0i}_{L}(\bar{k}) = 0,
\eeq
and
\beq
P^{ij}_{T}(\bar{k}) = \delta^{ij}-\frac{k^i k^j}{|\k|^2} = P^{ij}_{T}.
\eeq
Hence we may identify $P^{\mu \nu}_{T}$ with $u^{\mu} u^{\nu} + P^{\mu \nu}_{T}(\bar{k})$ (with $\tilde{P}^{\mu\nu}$ with $\tilde{P}^{\mu \nu}(k)$).
We write the full polarization as
\beq
\Pi^{\mu \nu}_{R} = \tilde{\Pi} \tilde{P}^{\mu \nu} + \Pi \tilde{P}^{\mu \nu}_{T}.
\eeq
The non-vanishing terms are
\beq
\Pi^{00}_{R} = \Pi \frac{|\k|^{2}}{(k^{0})^{2} - |\k|^{2}},
\eeq
\beq
\Pi^{xx}_R = \Pi \left(1 + \frac{(k^{x})^2}{(k^{0})^{2} - |\k|^2}\right)+\tilde{\Pi} \left(1 - \frac{(k^{x})^{2}}{|\k|^2} \right),
\eeq
and
\beq
\Pi^{yy}_R = \Pi \left(1 + \frac{(k^{y})^{2}}{(k^{0})^{2} - |\k|^{2}}\right)+\tilde{\Pi} \left(1 - \frac{(k^2)^2}{|\k|^2} \right).
\eeq

For kinematics where $\vec{k} = (|\k|,0)$, these expressions simplify to
\beq
\Pi^{00}_R =  \Pi \frac{|\k|^2}{(k^{0})^{2} - |\k|^{2}},
\eeq
\beq
\Pi^{xx}_R = \Pi \frac{(k^{0})^{2}}{(k^{0})^{2} - |\k|^{2}},
\eeq
and
\beq
\Pi^{yy}_R = \Pi + \tilde{\Pi}.
\eeq
Hence $\Pi^{xx}_{R}$ and $\Pi^{yy}_{R}$ are sufficient to determine the two functions $\Pi$ and $\tilde{\Pi}$.
\subsection{Imaginary part, full expressions}
The first contribution to both $\Pi^{xx}_{R}$ and $\Pi^{yy}_{R}$ is the constant term,
\beq
 2 \int_{\p}  \frac{1- 2n_F(E_{\p})}{2 E_{\p}} = -\frac{T \log 2}{\pi} + \Pi_{T=0}. 
\eeq
where $\Pi_{T=0}$ is the zero temperature contribution. Because this contribution is pure real, it drops out of the imaginary part.

Thus the full imaginary part of $\Pi^{ij}_R$ can be obtained from
\beq
\Im \Pi^{ij}_{R} = \Im \sum_{s,s'} \int_{\p} \frac{L^{ij}}{4 E_{\p} E_{\k-\p}}  \frac{s s' \left[ n_{F}(sE_{\p}) + n_{F}(s' E_{\k-\p})-1 \right]}{\omega + i \epsilon - s E_{\p} - s' E_{\k-\p}},\label{eq:polarizationexpression}
\eeq
where the numerator is
\beq
L^{ij} = - (-(k^{0})^{2} + |\k|^2) \delta^{ij} + 2 k^{i} k^{j} - 4 p^{i} p^{j}.
\eeq
Using kinematics where $\vec{k}=(|\k|,0)$ and $\vec{p}= (|\p| \cos \theta, |\p| \sin \theta)$, we find
\beq
L^{xx}= (k^{0})^{2} + |\k|^{2} - 4 |\p|^{2} \cos^{2} \theta
\eeq
and
\beq
L^{yy} = (k^{0})^{2} - |\k|^{2} - 4 |\p| \sin^2 \theta.
\eeq

We can check that $\Im \Pi^{ij}_R$ is an odd function of frequency. For positive $k^{0}$, only the $(1,1)$, $(1,-1)$, and $(-1,1)$ terms contribute. Furthermore, the $(1,-1)$ and $(-1,1)$ terms give the same result. The $(1,1)$ contribution is
\beq
\Im \Pi^{ij}_{R}\bigg|_{ss'=1} =- \int_p \frac{\pi L^{ij}}{4 E_{\p} E_{\k-\p}} \left[n_{F}(E_{\p})+ n_{F}(E_{\k-\p}) -1 \right]\delta(k^{0} - E_{\p} - E_{\k-\p}).
\eeq
The $(1,-1)$ contribution (counted twice) is
\beq
\Im \Pi^{ij}_R\bigg|_{ss'=-1} = 2 \int_{\p} \frac{\pi L^{ij}}{4 E_{\p} E_{\p-\k}} \left[n_{F}(E_{\p}) - n_{F}(E_{\k-\p}) \right]\delta(k^{0}- E_{\p}+E_{\k-\p}).
\eeq

The delta function condition (with $s=1$) is
\beq
k^{0} - E_{\p} = s' E_{\k-\p}.
\eeq
To solve it, let us square both sides,
\beq
(k^{0} - |\p|)^2 = (|\k|-|\p|\cos \theta)^{2} + |\p|^{2} \sin^{2}\theta.
\eeq
The result is
\beq
(k^{0})^{2} - 2k^{0}|\p| = |\k|^{2} - 2 |\p| |\k| \cos \theta
\eeq
or
\beq
|\p| = |\p|_0 = \frac{(k^{0})^{2} - |\k|^2}{2(k^{0} - |\k|\cos \theta)}.
\eeq
We need either $k^{0} > |\k|$ or $k^{0} < |\k| \cos \theta$ to have a positive solution for $|\p|$. The former is realized for $ss'=1$ while the latter is realized for $ss'=-1$. To complete the evaluation of the delta function, we also need
\beq
v_{s'}=\frac{\partial(E_{\p} + s' E_{\k-\p})}{\partial |\p|}\bigg|_{|\p|=|\p|_0} = 1 + s' \frac{|\p|-|\k|\cos \theta}{E_{\k-\p}}.
\eeq

The $ss'=1$ contribution is
\beq
\Im \Pi^{ij}_R\bigg|_{ss'=1} = -\frac{1}{(2\pi)^2} \int_0^{2\pi} d\theta \int_{0}^\infty d|\p| |\p| \frac{\delta(|\p|-|\p|_0)}{|v_1|} \frac{\pi L^{ij}\left(n_{F}(E_{\p})+ n_{F}(E_{\k-\p}) -1 \right)}{4 E_{\p} E_{\k-\p}} .
\eeq
The $ss'=-1$ contribution is
\beq
\Im \Pi^{ij}_{R}\Bigg|_{ss'=-1} = \frac{1}{(2\pi)^2} \int_{-\cos^{-1} \frac{k^{0}}{|\k|}}^{\cos^{-1} \frac{k^{0}}{|\k|}} d\theta \int_{0}^{\infty} d|\p| |\p| \frac{\delta(|\p|-|\p|_0)}{|v_{-1}|} \frac{\pi L^{ij} \left(n_{F}(E_{\p}) - n_{F}(E_{\k-\p}) \right)}{4 E_{\p} E_{\k-\p}}.
\eeq

The real part is then obtained from a subtracted Kramers-Kronig relation. For a function $\chi(\omega) = \chi_1 + i \chi_2$ obeying the assumptions of Kramers-Kronig, the real part in terms of the imaginary part is
\beq
\chi_1(\omega) = \frac{2}{\pi}\int_0^\infty d\nu \frac{\omega \chi_2(\omega) - \nu \chi_2(\nu)}{\omega^2 - \nu^2}.\label{eq:kramerskronig}
\eeq

\subsection{Low temperature limit}

Suppose $\omega \gg T$ and $|k|=0$. Only the $ss'=1$ term contributes to $\Im \Pi^{ij}_R$. The contribution is
\beq
\Im \Pi^{ij}_{R} = \frac{1}{(2\pi)^2} \int d|\p| |\p| \frac{\delta(|\p|-k^{0}/2)}{2} \frac{\pi L^{ij}}{4 |\p|^2}
\eeq
with $L^{xx}\rightarrow (k^{0})^2 - 4 |\p|^{2} \cos^{2}\theta$ and $L^{yy}\rightarrow (k^{0})^2 - 4 |\p|^{2} \sin^{2} \theta$. Because of the simple dependence on $\theta$, we conclude the $\Im \Pi^{xx}_{R} = \Im \Pi^{yy}_{R}$ in this limit. (Physically, since $|\k| =0$, there is no difference between longitudinal and transverse.)

We find
\beq
\Im \Pi^{yy}_{R} = \frac{1}{16\pi k^{0}} \int d\theta ((k^{0})^{2} - (k^{0})^{2} \sin^{2}\theta) = \frac{k^{0}}{16}.
\eeq

\subsection{High temperature limit}

In the high temperature limit, $k/T \ll 1$, we neglect $|\k|$ and $k^{0}$ relative to $T$ and the loop momentum whenever the dependence on $k$ is non-singular. The numerators limit to
\beq
L^{xx} = - 4 |\p|^{2} \cos^{2} \theta
\eeq
and
\beq
L^{yy}= - 4 |\p|^{2} \sin^{2} \theta.
\eeq

The $xx$ polarization is
\beq
\Pi^{xx}_{R} = \left( -\frac{T \log 2}{\pi} + \sum_{s,s'} \int_{\p} \frac{L^{xx}}{4 |\p|^{2}} \frac{s s' \left[ n_F(sE_{\p}) + n_{F}(s' E_{\k-\p})-1 \right]}{k^{0}+i\epsilon - s E_{\p} - s' E_{\k-\p}}\right).
\eeq
Both $ss'=1$ terms inside the parenthesis contribute the same value (in the small $k/T$ limit, both are real), as do both $ss'=-1$ terms.  These are
\beq
ss'=1: 2 \int_{\p} \frac{-4 \cos^{2} \theta}{4} \frac{2 n_{F}(|\p|)-1}{-2 |\p|} =  \frac{T\log 2}{2\pi} + \text{($T=0$ value)}
\eeq
and
\beq
ss'=-1: 2 \int_{\p} \frac{-4 \cos^{2} \theta}{4} \frac{-\frac{\partial n_{F}}{\partial |\p|}|\k| \cos \theta}{k^{0} +i\epsilon - |\k| \cos \theta} = -\frac{T \log 2}{2\pi^{2}}\int d\theta \cos^{2} \theta \frac{\cos \theta}{z-\cos \theta},
\eeq
where $z=(k^{0}+i\epsilon)/|k|$. Further processing the $ss'=-1$ term gives
\beq
ss'=-1: - \frac{T \log 2}{2\pi^{2}} \left( -\pi + \int \dm\theta \cos^2 \theta \frac{z}{z-\cos\theta}\right).
\eeq
Then we see that the constant term, the $ss'=1$ terms, and the constant part of the $ss'=-1$ terms combine to give zero. Thus the $xx$ polarization is
\beq
\Pi^{xx}_R = -\frac{T \log 2}{2\pi^2} \int d\theta \frac{z \cos^2 \theta}{z-\cos \theta}.
\eeq

The only difference between $xx$ and $yy$ is the replacement $\cos^2 \theta \rightarrow \sin^2 \theta$ when going from $L^{xx}$ to $L^{yy}$. Hence it follows that
\beq
\Pi^{yy}_R = -\frac{T \log 2}{2\pi^2} \int d\theta \frac{z \sin^2 \theta}{z-\cos \theta}.
\eeq

The two integrals we must calculate are
\beq
I_{1}=\int d\theta \frac{z \cos^{2} \theta}{z-\cos \theta}
\eeq
and
\beq
I_{2} = \int d\theta \frac{z \sin^{2} \theta}{z-\cos \theta}.
\eeq
The sum of $I_{1}$ and $I_{2}$ is
\beq
I_{1} + I_{2} =- \int \frac{dw}{i w} \frac{2wz}{w^2-2wz+1},
\eeq
where we converted the expression to a contour integral around the unit circle. The integrand has poles at $w=w_{\pm} = z\pm \sqrt{z^{2}-1}$. The $w=w_{-}$ pole is inside the unit circle for $\Im z >0$. The integral is thus
\beq
I_{1} + I_{2} = - 2\pi \frac{2z}{w_{-} - w_{+}} = \frac{2\pi z}{\sqrt{z^{2}-1}}.
\eeq

The difference of $I_{1}$ and $I_{2}$ is
\beq
I_{1} - I_{2} = - \int \frac{d w}{i w} \frac{2wz}{w^{2}-2wz+1} \frac{w^{4}+1}{2 w^{2}},
\eeq
where again we converted to a contour integral. The integrand now has poles at $w=0$ (double pole) and at $w=w_{\pm} = z \pm \sqrt{z^{2}-1}$. The integral is thus
\beq
I_{1}-I_{2} = - 2\pi \left( 2 z^{2} +\frac{2z}{w_{-} - w_{+}}\frac{w_{-}^{4}+1}{2w_-^{2}} \right).
\eeq
Simplifying gives
\beq
I_1 - I_{2} = - 2\pi \left( 2 z^{2} - \frac{z}{\sqrt{z^{2}-1}} \left( \frac{(w_{-}^{2}+1)^2}{2w_{-}^{2}} - 1\right)\right),
\eeq
or using $w_{-}^{2}+1 = 2 z w_{-}$,
\beq
I_{1} - I_{2} = 2\pi \left(  \frac{z}{\sqrt{z^{2}-1}}(2 z^{2}-1) - 2 z^{2} \right).
\eeq

Hence
\beq
\Pi^{xx}_{R} = -\frac{T \log 2}{2 \pi}\left[\frac{z}{\sqrt{z^{2}-1}} + \frac{z}{\sqrt{z^{2}-1}}(2z^{2}-1) - 2 z^{2} \right]
\eeq
or
\beq
\Pi^{xx}_{R} = -\frac{T \log 2}{2 \pi}\left[\frac{2z^{3}}{\sqrt{z^{2}-1}} - 2 z^{2} \right] .
\eeq
Similarly,
\beq
\Pi^{yy}_{R} = -\frac{T \log 2}{2 \pi}\left[\frac{z}{\sqrt{z^{2}-1}} - \frac{z}{\sqrt{z^{2}-1}}(2z^{2}-1) + 2 z^{2} \right]
\eeq
or
\beq
\Pi^{yy}_{R} = -\frac{T \log 2}{2 \pi}\left[\frac{z(2-2z^{2})}{\sqrt{z^{2}-1}}  + 2 z^{2} \right].
\eeq

\section{Fermion damping rate}
\label{appendix:damping}
We can verify the expression for the fermion damping directly from the optical theorem using the finite temperature cutting rules  \cite{Weldon:1983jn}. This relates the decay rate to the imaginary part of the tree level cross-section which in turn is related to the imaginary part of the self-energy. We can define $\mathcal{M}_{L}$ and $\mathcal{M}_{T}$ as the longitudinal and transverse matrix elements for the tree level scattering processes of an electron emitting a soft photon
\begin{equation}
\mathcal{M}_{L(T)}=\frac{1}{\sqrt{N_{F}}}\bar{u}(p)(-i\gamma_{\mu})u(k)\epsilon^{\mu}_{L(T)}(q^{0},\k-\p) 
\end{equation}
where $\bar{u}(p)$ and $u(k)$ are massless on shell spinors.
We can write $\Gamma_{\k}$ in terms of these matrix elements as 
\begin{eqnarray}
\Gamma_{\k}&\equiv&\frac{1}{2E_{\k}}\int_{\p}\int\frac{\mathrm{d}p^{0}}{2\pi}\frac{\mathrm{d}q^{0}}{2\pi}\sum_{\lambda=L,T}\mathcal{A}_{\lambda}(q^{0},\k-\p)2\pi \delta((p^{0})^{2}-E^{2}_{\p})\vert\mathcal{M}_{\lambda}\vert^{2}(1-n_{F}(p^{0})+n_{B}(q^{0}))
\nn&&2\pi\delta(k^{0}-p^{0}-q^{0})
\nn
&=&\int_{\p}\int\frac{\mathrm{d}q^{0}}{2\pi}\sum_{s=\pm}\sum_{\lambda=L,T}\frac{s}{8E_{\k}E_{\p}}\mathcal{A}_{\lambda}(q^{0},\k-\p)\vert\mathcal{M}_{\lambda}\vert^{2}
(1-n_{F}(sE_{\p})+n_{B}(q^{0}))2\pi\delta(k^{0}-sE_{\p}-q^{0})
\nn
&=&\sum_{\lambda=L,T}\sum_{s=\pm}\int_{\p}\frac{s}{8E_{\k}E_{\p}}
\mathcal{A}_{\lambda}(k-p_{s})\vert\mathcal{M}_{\lambda}\vert^{2}
(1-n_{F}(s E_{\p})+n_{B}(k^{0}-s E_{\p}))
\label{eq:decaycross}
\end{eqnarray}
\begin{figure}[t]
	\centering
	\includegraphics[width=0.5\textwidth]{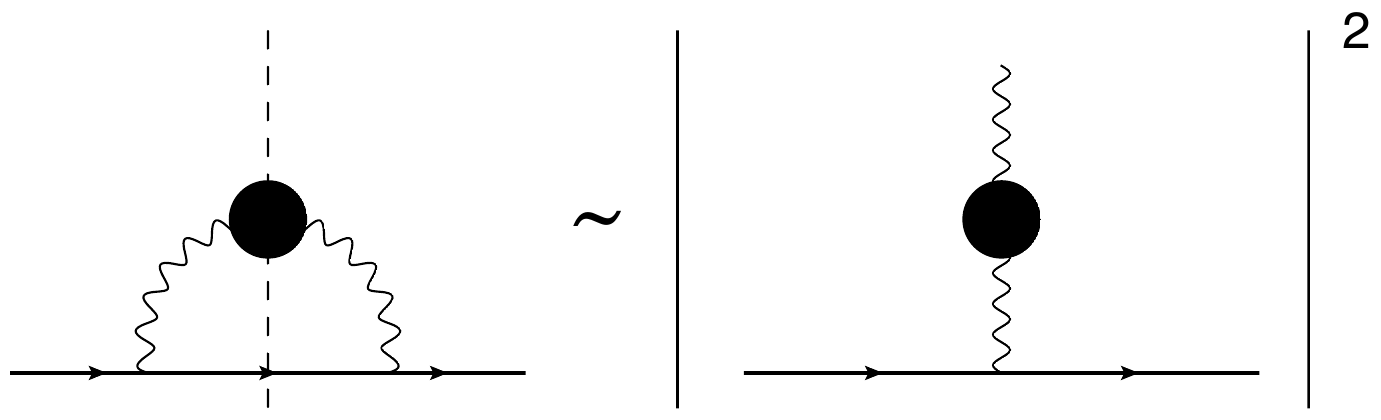}\label{fig:fermiondecay}
	\caption{The decay rate is related to the imaginary part of the fermion self energy via the optical theorem.}
\end{figure}
where $k=(k^{0},\k)$ and $p_{s}=(sE_{\p},\p)$. Plugging $\mathcal{M}_{L(T)}$ into \ref{eq:decaycross} and using the relation $u(k)\bar{u}(k)=\slashed{k}$ we get
\begin{eqnarray}
\Gamma_{\k}&=&\sum_{\lambda=L,T}\sum_{s=\pm}\int_{\p}\frac{\text{Tr}\left[s\bar{u}(k)\gamma_{\mu}\slashed{p}_{s}\gamma_{\nu}u(k)\epsilon^{\mu}_{\lambda}(\k-\p)\epsilon^{\nu}_{\lambda}(\k-\p) \vert\mathcal{M}_{\lambda}\vert^{2}\right]}{8E_{\k}E_{\p}}
\mathcal{A}_{\lambda}(k-p_{s})
\nn
&&(1-n_{F}(s E_{\p})+n_{B}(k^{0}-s E_{\p}))
\nn &=&
 -\frac{1}{2E_{\k}}\text{Tr}\left[\bar{u}(k)\Im \Sigma_{R}(\slashed{k})u(k)\right]\Big\vert_{k^{0}=E_{\k}}
 \nn &=&-\frac{1}{2E_{\k}}\Im\text{Tr}[\slashed{k}\mathrm{\Sigma}_{R}(\slashed{k})]\Big\vert_{k^{0}=E_{\k}}
\end{eqnarray}
we find the  relation for the self energy and the decay rate given in Eqn.~\ref{eq:decay}. We can now write the decay rate $\Gamma_{\k}$ in Coulomb gauge as
\begin{eqnarray}
\Gamma_{\k}
&=&
\frac{1}{N_{F}}\sum_{s=\pm}\int_{\p}\frac{s}{4E_{\k}E_{\p}}
\Big(\mathcal{A}_{L}(k-p_{s})
(2(k\cdot u)(p_{s}\cdot u)-k\cdot p_{s}u\cdot u)
\nn &+&\mathcal{A}_{T}(k-p_{s})
(2(k\cdot\epsilon_{T}(\k-\p))(p_{s}\cdot\epsilon_{T}(\k-\p))-k\cdot p_{s}(\epsilon_{T}(\k-\p)\cdot \epsilon_{T}(\k-\p))\Big)
\nn
&&(1-n_{F}(s E_{\p})+n_{B}(k^{0}-s E_{\p}))\Bigg\vert_{k^{0}=E_{\k}}\label{eq:fulldecayapp}
\end{eqnarray}
Defining the Lorentz vectors $k=(k^{0},\vert\k\vert,0)$, $p_{s}=(sE_{\p},\vert\p\vert\cos\theta,\vert\p\vert\sin\theta)$, and the transverse polarization vector $\epsilon^{\mu}_{T}(\k-\p)=\frac{1}{\vert\k-\p\vert}\left(0, \vert\p\vert\sin\theta,
\vert\k\vert-\p\cos\theta\right)$, we get the final expression for $\Gamma_{\k}$ in Coulomb gauge
\begin{eqnarray}
\label{eq:fulldecaycoloumb}
\Gamma_{\k}&=&\frac{1}{N_{F}}\int_{\p}\sum_{s=\pm}\frac{s}{4E_{\k}E_{\p}}\Bigg[\Bigg(\mathcal{A}_{L}(k-\tilde{p}_{s})
\left(sE_{\k}E_{\p}+\vert \k\vert\vert \p\vert\cos\theta\right)
\nn
&+&\mathcal{A}_{T}(k-\tilde{p}_{s})
\Big(s E_{\k}E_{\p}-\vert\k\vert\vert\p\vert\cos\theta+\frac{2\vert\k\vert^{2}\vert\p\vert^{2}(1-\cos^{2}\theta)}{\vert\k\vert^{2}+\vert\p\vert^{2}-2\vert\k\vert\vert\p\vert\cos\theta}\Big)
\Bigg)(1-n_{f}(s E_{\p})+n_{b}(E_{\k}-s E_{\p}))
\Bigg].\label{eq:finalgamma}
\end{eqnarray}

\section{Full expressions for the scalar kernel}
 \label{appendix:kernel}
Using the definition in Eq. \ref{eq:scalarkernel}, we can write an expression for the scalar kernel in the Bethe-Salpeter equation in terms of the rung matrix elements. 
\begin{eqnarray}
K_{ss^{\prime}}(\k,\k^{\prime})&=&e^{2i(\theta_{\k^{\prime}}-\theta_{\k})}\mathcal{R}_{11}-e^{i(\theta_{\k^{\prime}}-\theta_{\k})}s_{1}s_{2}(\mathcal{R}_{22}+\mathcal{R}_{23}+\mathcal{R}_{32}+\mathcal{R}_{33})+\mathcal{R}_{44}
\nn
&-&e^{-2i\theta_{\k}}\mathcal{R}_{14}
-e^{2i\theta_{\k^{\prime}}}\mathcal{R}_{41}+ie^{i(2\theta_{\k^{\prime}}-\theta_{\k})}s_{1}(\mathcal{R}_{21}+\mathcal{R}_{31})+ie^{i(\theta_{\k^{\prime}}-2\theta_{\k})}s_{2}(\mathcal{R}_{12}+\mathcal{R}_{13})
\nn&-&ie^{i\theta_{\k^{\prime}}}s_{2}(\mathcal{R}_{42}+\mathcal{R}_{43})-ie^{-i\theta_{\k}}s_{1}(\mathcal{R}_{24}+\mathcal{R}_{34})
\end{eqnarray}
where $\theta_{\k}$ is the angle between $\k$ and the $x$ axis and $\theta_{\k^{\prime}}$ is the angle between $\k^{\prime}$ and the $x$ axis.
\\
If we make a rotationally symmetric ansatz for the wavefunction, only terms depending on $\theta_{\k^{\prime}}-\theta_{\k}$ remain, which leaves us
\begin{eqnarray}
K_{ss^{\prime}}(\k,\k^{\prime})&=&e^{2i(\theta_{\k^{\prime}}-\theta_{\k})}\mathcal{R}_{11}-e^{i(\theta_{\k^{\prime}}-\theta_{\k})}s_{1}s_{2}(\mathcal{R}_{22}+\mathcal{R}_{23}+\mathcal{R}_{32}+\mathcal{R}_{33})+\mathcal{R}_{44}\label{eq:kernelcompute}
\end{eqnarray}
Both rungs can be written in the form 
\begin{eqnarray}
\mathcal{R}=\mathcal{R}_{0,0}\gamma_{0}\otimes\gamma_{0}+\mathcal{R}_{i,j}\gamma_{i}\otimes\gamma_{j}+\mathcal{R}_{i,0}\gamma_{i}\otimes\gamma_{0}+\mathcal{R}_{0,j}\gamma_{0}\otimes\gamma_{j}
\end{eqnarray}
For the first rung, we can write these coefficients as 
\begin{eqnarray}
\mathcal{R}_{1}(\bar{k})_{(0,0)}&=&-\frac{\mathcal{Q}_{b}(\bar{k}^{0})}{N_{F}}\mathcal{A}_{L}(\bar{k})
\nn
\mathcal{R}_{1}(\bar{k})_{(x,x)}&=&-\frac{\mathcal{Q}_{b}(\bar{k}^{0})}{N_{F}}\mathcal{A}_{T}(\bar{k})\left(1-\frac{(\vert\k\vert-|\k^{\prime}|\cos\theta_{\k,\k^{\prime}})^{2}}{\vert\bar{\k}\vert^{2}})\right)
\nn
\mathcal{R}_{1}(\bar{k})_{(y,y)}&=&-\frac{\mathcal{Q}_{b}(\bar{k}^{0})}{N_{F}}\mathcal{A}_{T}(\bar{k})\left(1-\frac{(\vert\k^{\prime}\vert\sin\theta_{\k,\k^{\prime}})^{2}}{\vert\bar{\k}\vert^{2}})\right)
\nn
\mathcal{R}_{1}(\bar{k})_{(x,y)}&=&-\frac{\mathcal{Q}_{b}(\bar{k}^{0})}{N_{F}}\mathcal{A}_{T}(\bar{k})\left(\frac{(\vert\k\vert-|\k^{\prime}|\cos\theta_{\k,\k^{\prime}})(\vert\k^{\prime}\vert\sin\theta_{\k,\k^{\prime}})}{\vert\bar{\k}\vert^{2}})\right)
\nn
\mathcal{R}_{1}(\bar{k})_{(y,x)}&=&-\frac{\mathcal{Q}_{b}(\bar{k}^{0})}{N_{F}}(\mathcal{A}_{T}(\bar{k})\left(\frac{(\vert\k\vert-|\k^{\prime}|\cos\theta_{\k,\k^{\prime}})(\vert\k^{\prime}\vert\sin\theta_{\k,\k^{\prime}})}{\vert\bar{\k}\vert^{2}})\right)
\end{eqnarray}
where $\theta_{\k,\k^{\prime}}$ is the angle between $\k$ and $\k^{\prime}$.
For the second rung, we use the general expression given in \ref{eq:numeratorcoefficients}.
Defining $\theta_{\p}$ as the angle between $\p$ and $\bar{\k}$ and $\phi$ as the angle between $\bar{\k}$ and $\K$, the transverse polarization vector with respect to the $\bar{\k}$ axis can be written as
\begin{eqnarray}
\epsilon^{\mu}_{T}(\p+\K)&=&\frac{1}{\vert\p+\K\vert}\left(0,-(\vert\p\vert\sin\theta_{\p}+\vert\K\vert\sin\phi),
\vert\p\vert\cos\theta_{\p}+\vert\K\vert\cos\phi \right)
\label{eq:transversepolarization}
\end{eqnarray}
We use these to define the following Lorentz vectors which are contained in the numerators of the longitudinal and transverse components of the second rung in 
\begin{subequations}
    \begin{align}
    \mathcal{R}^{(1)}_{2,L}(\bar{k},K,\p)^{\mu}&=\left(\bar{k}^{0}-s^{\prime}E_{-}, -\bar{\k}/2-\p\right) 
    \\
    \mathcal{R}^{(2)}_{2,L}(\bar{k},K,\p)^{\mu}&=\left(s^{\prime}E_{-}, -\bar{\k}/2+\p\right)
    \\
\mathcal{R}^{(1)}_{2,T}(\bar{k},K,\p)^{\mu}&=\left(\bar{k}^{0}-s^{\prime}E_{-}, (\bar{\k}+2\p)\cdot\epsilon_{T}(\p+\K)\epsilon_{T}(-\p-\K)+\bar{\k}/2+\p\right)
\\
\mathcal{R}^{(2)}_{2,T}(\bar{k},K,\p)^{\mu}&=\left(s^{\prime}E_{-}, (\bar{\k}-2\p)\cdot\epsilon_{T}(\p+\K)\epsilon_{T}(-\p-\K)+\bar{\k}/2-\p\right)
    \end{align}
    \label{eq:fourvectornumerators}
\end{subequations}
Plugging \ref{eq:transversepolarization} into \ref{eq:fourvectornumerators}, we  can write the spatial components of the Lorentz vectors for the longitudinal polarizations  as
\begin{subequations}
    \begin{align}
    \mathcal{R}^{(1)}_{2,L}(\bar{k},K,\p)^{x}&= -\vert\bar{\k}\vert/2-\vert\p\vert\cos\theta_{\p} 
    \\
     \mathcal{R}^{(1)}_{2,L}(\bar{k},K,\p)^{y}&= -\vert\p\vert\sin\theta_{\p}
    \\
    \mathcal{R}^{(2)}_{2,L}(\bar{k},K,\p)^{x}&= -\vert\bar{\k}\vert/2+\vert\p\vert\cos\theta_{\p}
    \\
    \mathcal{R}^{(2)}_{2,L}(\bar{k},K,\p)^{y}&=\vert\p\vert\sin\theta_{\p}
    \end{align}
       \label{eq:fourvectornumeratorsl}
    \end{subequations}
and the transverse polarizations as
\begin{subequations}
\begin{align}
\mathcal{R}^{(1)}_{2,T}(\bar{k},K,\p)^{x}&=
\frac{(2\vert\K\vert\vert\p\vert\sin(\theta_{\p}-\phi)-\vert\k\vert(\vert\p\vert\sin\theta_{\p}+\vert\K\vert\sin\phi))(\vert\p\vert\sin\theta_{\p}+\vert\K\vert\sin\phi)}{\vert\p+\K\vert^{2}}
+\frac{\vert\bar{\k}\vert+2\vert\p\vert\cos\theta_{\p}}{2}
\\
\mathcal{R}^{(1)}_{2,T}(\bar{k},K,\p)^{y}&=-\frac{(2\vert\K\vert\vert\p\vert\sin(\theta_{\p}-\phi)-\vert\k\vert(\vert\p\vert\sin\theta_{\p}+\vert\K\vert\sin\phi))(\vert\p\vert\cos\theta_{\p}+\vert\K\vert\cos\phi)}{\vert\p+\K\vert^{2}}+\p\sin\theta_{\p}
\\
\mathcal{R}^{(2)}_{2,T}(\bar{k},K,\p)^{x}&= \frac{(2\vert\K\vert\vert\p\vert\sin(\theta_{\p}-\phi)+\vert\k\vert(\vert\p\vert\sin\theta_{\p}\vert\K\vert\sin\phi))(\vert\p\vert\sin\theta_{\p}+\vert\K\vert\sin\phi)}{\vert\p+\K\vert^{2}}+\frac{\vert\bar{\k}\vert-2\vert\p\vert\cos\theta_{\p}}{2}
\\
\mathcal{R}^{(2)}_{2,T}(\bar{k},K,\p)^{y}&=-\frac{(2\vert\K\vert\vert\p\vert\sin(\theta_{\p}-\phi)+\vert\k\vert(\vert\p\vert\sin\theta_{\p}+\vert\K\vert\sin\phi))(\vert\p\vert\cos\theta_{\p}+\vert\K\vert\cos\phi)}{\vert\p+\K\vert^{2}}-\vert\p\vert\sin\theta_{\p}
    \end{align}
    \label{eq:fourvectornumeratorst}
\end{subequations} 
We can now now plug the expressions in \ref{eq:fourvectornumeratorst} into \ref{eq:numeratorcoefficients} and evaluate the momentum integrals to get complete expressions for the matrix coefficients of rung II.
\section{Feynman rules for ladder diagrams}
We derive the Feynman rules for the ladder diagrams in the $\frac{1}{N_{F}}$ expansion. Ignoring thermal corrections, we can start by going into the interaction picture with respect to the the free Hamiltonian. The time evolution operator is given by
\begin{equation}
U_{I}=\mathcal{T}\text{exp}\left(-\frac{1}{\sqrt{N_{F}}}\sum^{N_{F}}_{a=1}\int^{t}_{0}\dm s\int_{\x}A_{\mu}(s,\x)j_{a}^{\mu}(s,\x)\right)
\end{equation}
here $A$ is the free photon field and $j_{\mu}$ is the free fermion current $j^{\mu}(t)=\bar{\psi}(t)\gamma^{\mu}\psi(t)$. We can expand this (dropping spatial arguments) as
\begin{eqnarray}
U_{I}=1+\frac{i}{2}\int^{t}_{0}\dm s_{1}A^{\mu}(s_{1})j_{\mu}(s)+\left(\frac{i}{2}\right)^{2}\int^{t}_{0}\dm s_{1}\int^{s_{1}}_{0}\dm s_{2}A^{\mu}(s_{1})j_{\mu}(s_{1})A^{\mu}(s_{2})j_{\mu}(s_{2})+...
\end{eqnarray}
for $U^{\dagger}_{I}$ we have
\begin{eqnarray}
U^{\dagger}_{I}=1-\frac{i}{2}\int^{t}_{0}\dm s_{1}A^{\mu}(s_{1})j_{\mu}(s_{1})+\left(\frac{-i}{2}\right)^{2}\int^{t}_{0}\dm s_{1}\int^{s_{1}}_{0}\dm s_{2}A^{\mu}(s_{2})j_{\mu}(s_{2})A^{\mu}(s_{1})j_{\mu}(s_{1})
\nn &+&....
\end{eqnarray}
We start by stating the usual anti-commutation relations for free fermions
\begin{subequations}
	\begin{align}
	\lbrace \psi_{a}(t,\x),\psi_{b}(0)\rbrace&=0
	\\
	\lbrace \bar{\psi}_{a}(t,\x),\bar{\psi}_{b}(0)\rbrace&=0
	\\
	\lbrace \psi_{a}(t,\x),\bar{\psi}_{b}(0)\rbrace&=\gamma^{0}\delta_{ab}\delta(t)\delta(\x).
	\end{align}
\end{subequations}
For currents we have
\begin{subequations}
	\begin{align}
		[\psi_{a}(t,\x),j^{\mu}_{b}(0)]&=\lbrace \psi_{a}(t,\x),\bar{\psi}_{b}(0)\rbrace \gamma^{\mu}\psi_{b}(0)
		\\
			[\bar{\psi}_{a}(t,\x),j^{\mu}_{b}(0)]&=-\bar{\psi}_{b}(0)\gamma^{\mu}\lbrace \bar{\psi}_{a}(t,\x),\psi_{b}(0)\rbrace
\\
	[j^{\mu}_{a}(t,\x),j^{\nu}_{b}(0)]&=\bar{\psi}_{a}(t,\x)\gamma^{\mu}\lbrace \psi(t,\x),\bar{\psi}_{b}(0)\rbrace\gamma^{\nu}\psi_{b}(0)
	\nn
	&-\bar{\psi}_{b}(0)\gamma^{\nu}\lbrace \bar{\psi}_{a}(t,\x),\psi_{b}(0)\rbrace\gamma^{\mu}\psi_{a}(t,\x).
	\end{align}
\end{subequations}
The current-fermion and current-current commutators can be written in terms of the retarded propagator terms of the retarded propagator is now
\begin{subequations}
\begin{align}
[\psi_{a}(t,\x),j^{\mu}_{b}(0)]&=G_{R,ab}(t,\x) \gamma^{\mu}\psi_{b}(0)
\\
[\bar{\psi}_{a}(t,\x),j^{\mu}_{b}(0)]&=-\bar{\psi}_{b}(0)\gamma^{\mu}G^{\ast}_{R,ab}(t,\x)
\\
[j^{\mu}_{a}(t,\x),j^{\nu}_{b}(0)]&=\bar{\psi}_{a}(t,\x)\gamma^{\mu} G_{R,ab}(t,\x)\gamma^{\nu}\psi_{b}(0)
-\bar{\psi}_{b}(0)\gamma^{\nu}G^{\ast}_{R,ab}(t,\x)\gamma^{\mu}\psi_{a}(t,\x). \label{eq:jjretarded}
\end{align}
\end{subequations}
Lastly for free fields, both $j^{\mu}_{a}(t,\x)$ and $\tilde{\psi}(t,\x)$ commute with the photon operator so we have
\begin{eqnarray}
[j^{\mu}_{a}(t,\x),A^{\nu}(0)]=[A^{\mu}_{i}(t,\x),\psi(0)]=[A^{\mu}_{i}(t,\x),\bar{\psi}(0)]=0.
\end{eqnarray}

\subsection{Fermion Ladder}
\label{appendix:ladders}
In the interaction picture, $\mathcal{F}(t,\x)$ is given by
\begin{equation}
\mathcal{F}(t,\x)=\frac{1}{N^{2}_{F}}\sum^{N_{F}}_{i,j=1}\text{Tr}\Big[\sqrt{\hat{\rho}}\lbrace U^{\dagger}_{I}\psi_{i}(t,\x)U_{I},\bar{\psi}_{j}(0)\rbrace\sqrt{\hat{\rho}}\lbrace U^{\dagger}_{I}\psi_{i}(t,\x)U_{I},\bar{\psi}_{j}(0)\rbrace^{\dagger}\Big]
\end{equation}
where $U^{\dagger}_{I}\tilde{\psi}(t)U_{I}$ is given by the expansion
\begin{eqnarray}
U^{\dagger}_{I}\psi(t)U_{I}&=&\psi(t)+\frac{i}{2}\int^{t}_{0}\dm s_{1}[\psi(t),iA^{\rho}(s_{1})j_{\rho}(s_{1})]
\nn
&+&
\left(\frac{-i}{2}\right)^{2}\int^{t}_{0}\dm s_{1}\int^{s_{1}}_{0}\dm s_{2}[[\psi(t),iA^{\rho}
(s_{1})j_{\rho}	(s_{1})],iA^{\sigma}	(s_{2})j_{\sigma}(s_{2})]
\nn&+&...\label{eq:interactionpsi}
\end{eqnarray} 
From Eqn.~\ref{eq:interactionpsi} and using the definition of the retarded propagator, we can immediately see that the zeroth order term in $\mathcal{F}(t,\x)$ is given by 
\begin{equation}
\mathcal{F}_{0}(t,\x)=[iG_{R}(t,\x)][iG^{\ast}_{R}(t,\x)]
\end{equation}
The first nontrivial term is given by
\begin{eqnarray}
\mathcal{F}_{1}(t,\x)&\sim&\text{Tr}\Big[\sqrt{\hat{\rho}}\Big\lbrace\frac{i}{2}\int^{t}_{0}\dm s\int_{\y}[\psi(t,\x),iA^{\mu}(s,\y)j_{\mu}(s,\y)],\bar{\psi}(0)\Big\rbrace
\nn
&&\sqrt{\hat{\rho}}\Big\lbrace\frac{i}{2}\int^{t}_{0}\dm s^{\prime}\int_{\y^{\prime}}[\psi(t,\x),iA^{\nu}(s^{\prime},\y^{\prime})j_{\nu}(s^{\prime},\y^{\prime})],\bar{\psi}(0)\Big\rbrace^{\dagger}\Big]
\end{eqnarray}
Expanding the anti-commutator for each time fold, we get
\begin{eqnarray}
\lbrace[\psi(t,\x),iA^{\mu}(s,\y)j_{\mu}(s,\y)],\bar{\psi}(0)\rbrace
&=&
i\lbrace A^{\mu}(s,\y)[\psi(t,\x),j_{\mu}(s,\y)],\bar{\psi}(0)\rbrace
\nn
&=& iA^{\mu}(s,\y)\lbrace\lbrace \psi(t,\x),\bar{\psi}(s,\y)\rbrace\gamma_{\mu} \psi(s,\y),\bar{\psi}(0)\rbrace
\nn
&=&iA^{\mu}(s,\y)(iG_{R}(t-s,\x-\y))\gamma_{\mu}(iG_{R}(s,\y))
\end{eqnarray}
Putting everything together gives us
\begin{eqnarray}
\mathcal{F}_{1}(t,\x)&=&-\frac{1}{4N_{F}}\int_{ss^{\prime}}\mathcal{D}^{\mu\nu}_{W}(s-s^{\prime},\y-\y^{\prime})G_{R}(s,\y)\gamma_{\mu}G_{R}(t-s,\x-\y)
\nn
&\times&[G^{\ast}_{R}(s^{\prime},\y^{\prime})]\gamma_{\nu}[G^{\ast}_{R}(t-s^{\prime},\x-\y^{\prime})]
\end{eqnarray}
which we identify as a ladder with one type one rung.
\\
Now we can expand to second order on both time folds. We consider the expansion of the commutator contained inside the anti-commutator on one time fold
\begin{eqnarray}
[[,],]&=&[[\psi(t),iA^{\mu}(s_{1})j_{\mu}(s_{1})],
iA^{\nu}(s_{2})
j_{\nu}(s_{2})]
\nn
&=&
i^{2}\lbrace \psi(t),\bar{\psi}(s_{1})\rbrace \gamma_{\mu}\psi(s_{1})j_{\nu}(s_{2})
[A^{\mu}(s_{1}),A^{\nu}(s_{2})]
\nn&+&i^{2}
\lbrace \psi(t),\bar{\psi}(s_{1})\rbrace \gamma_{\mu}[\psi(s_{1}),j_{\nu}(s_{2})]
A^{\mu}(s_{1})
A^{\nu}(s_{2})
\end{eqnarray}
From the first term we get
\begin{eqnarray}
\lbrace[[,],],\rbrace_{1}&=&i^{2}\lbrace\lbrace \psi(t),\bar{\psi}(s_{1})\rbrace \gamma_{\mu}\psi(s_{1})j_{\nu}(s_{2})
[A^{\mu}(s_{1}),A^{\nu}(s_{2})],\bar{\psi}(0)\rbrace
\nn
&=&i^{2}\lbrace \psi(t),\bar{\psi}(s_{1})\rbrace\gamma_{\mu}\psi(s_{1})\bar{\psi}(s_{2})\gamma_{\nu}\lbrace\bar{\psi}(0),\psi(s_{2})\rbrace[A^{\mu}(s_{1}),A^{\nu}(s_{2})]
\nn
&=&i^{2}(iG_{R}(t-s_{1}))\gamma_{\mu}\psi(s_{1})\bar{\psi}(s_{2})\gamma_{\nu}(iG^{\ast}_{R}(s_{2})[A^{\mu}(s_{1}),A^{\nu}(s_{2})]
\end{eqnarray}
which will give us the type II rung. This can be written as
\begin{eqnarray}
\mathcal{F}_{2}(t,\x)&=&-\frac{1}{4N_{F}}\int_{s_{1}s^{\prime}_{1}}
\int_{s_{2}s^{\prime}_{2}}\mathcal{D}^{\mu\nu}(s_{1}-s_{2})\mathcal{D}^{\mu^{\prime}\nu^{\prime}}(s^{\prime}_{1}-s^{\prime}_{2})
\nn
&&G_{R}(t-s_{1}))\gamma_{\mu}G_{W}(s_{1}-s^{\prime}_{1})\gamma_{\mu^{\prime}}G^{\ast}_{W}(s_{2}-s^{\prime}_{2})\gamma_{\nu}G^{\ast}_{R}(s_{2})\gamma_{\nu^{\prime}}
\end{eqnarray}

The second term gives
\begin{eqnarray}
\lbrace[[,],]\rbrace_{2}&=&i^{2}\lbrace[[\psi(t),j_{\mu}(s_{1})],j_{\nu}(s_{2})]A^{\mu}(s_{1})
A^{\nu}(s_{2}),\bar{\psi}(0)\rbrace
\nn
&=&
i^{2}\lbrace \psi(t),\bar{\psi}(s_{1})\rbrace\gamma_{\mu}\lbrace \psi(s_{1}),\bar{\psi}(s_{2})\rbrace \gamma_{\nu}\lbrace\psi(s_{2}),\bar{\psi}(0)\rbrace A^{\mu}(s_{1})
A^{\nu}(s_{2})
\nn
&=&
i^{2}G_{R}(t-s_{1})\gamma_{\mu}G_{R}(s_{1}-s_{2}) \gamma_{\nu}G_{R}(s_{2})A^{\mu}(s_{1})A^{\nu}(s_{2})
\end{eqnarray}
This will give two copies of the type I rung or a self energy term for $\psi$.
\subsection{Squared Current-Current Commutator}
In thr interaction $\mathcal{C}(t,\x)$ is given by
\begin{equation}
\mathcal{C}(t,\x)=\frac{1}{N^{2}_{F}}\sum^{N_{F}}_{i,j=1}\delta_{\mu\rho}\delta_{\nu\sigma}\text{Tr}\Big[\sqrt{\hat{\rho}}\lbrace U^{\dagger}_{I}j^{\mu}_{i}(t,\x)U_{I},j^{\nu}_{j}(0)\rbrace\sqrt{\hat{\rho}}\lbrace U^{\dagger}_{I}j^{\rho}_{i}(t,\x)U_{I},j^{\sigma}_{j}(0)\rbrace^{\dagger}\Big]
\end{equation}
The zeroth order term can be deduced from the expression in \ref{eq:jjretarded}
\begin{eqnarray}
\mathcal{C}_{0}(t,\x)&=&G^{\ast}_{W}(t,\x)\gamma^{\mu}iG_{R}(t,\x)\gamma^{\nu}G_{W}(0)\gamma^{\nu} G^{\ast}_{R}(t,\x)\gamma^{\mu}]
\nn
&+&G^{\ast}_{W}(0)\gamma^{\nu}iG^{\ast}_{R}(t,\x)\gamma^{\mu}G_{W}(t,\x)\gamma^{\mu}iG_{R}(t,\x)\gamma^{\nu}
\nn
&-&G^{\ast}_{W}(t,\x)\gamma^{\mu} iG_{R}(t,\x)\gamma^{\nu}G^{\ast}_{W}(t,\x)\gamma^{\mu} iG_{R}(t,\x)\gamma^{\nu}
\nn
&-&G^{\ast}_{W}(t,\x)\gamma^{\nu} iG^{\ast}_{R}(t,\x)\gamma^{\mu}G^{\ast}_{W}(t,\x)\gamma^{\nu} iG^{\ast}_{R}(t,\x)\gamma^{\mu}
\end{eqnarray}
Diagrammatically the first terms each look like a box but have momentum flowing in the opposite directions. The second two terms also have opposite momentum flow but have Wightman functions which are crossed. However, ladders with crossed legs should not contribute to chaos given the pole structure \cite{Chowdhury:2017jzb}.
\\
The first nontrivial in the expansion for the current-current commutator is
\begin{eqnarray}
\mathcal{C}_{1}(t,\x)&=&\text{Tr}\Big[\sqrt{\hat{\rho}}\Big\lbrace\frac{i}{2}\int^{t}_{0}\dm s\int_{\y}[j^{\mu}(t,\x),iA^{\nu}(s,\y)j_{\nu}(s,\y)],j^{\rho}(0)\Big\rbrace
\nn
&&\sqrt{\hat{\rho}}\Big\lbrace\frac{i}{2}\int^{t}_{0}\dm s^{\prime}\int_{\y^{\prime}}[j^{\mu}(t,\x),iA^{\nu}(s^{\prime},\y^{\prime})j_{\nu}(s^{\prime},\y^{\prime})],j^{\rho}(0)\Big\rbrace^{\dagger}\Big]
\end{eqnarray}
Expanding the commutator on one of the time folds (again supressing spatial arguments) gives us
\begin{eqnarray}
[[j_{\mu}(t),A^{\nu}(s_{1})j_{\nu}(s_{1})],j_{\rho}(0)]&=&A^{\nu}(s_{1})[[j_{\mu}(t),j_{\nu}(s_{1})],j_{\rho}(0)]
\nn
&=&A^{\nu}(s_{1})\bar{\psi}(t)\gamma^{\mu}G_{R} (t-s_{1})\gamma^{\nu}G_{R}(s_{1})\gamma^{\rho}\psi(0)
\nn
&-&A^{\nu}(s_{1})\bar{\psi}(0)\gamma^{\rho}G^{\ast}_{R}(s_{1})\gamma^{\nu}G^{\ast}_{R}(t-s_{1})\gamma^{\mu}\psi(t)
\nn
&-&A^{\nu}(s_{1})\bar{\psi}(0)\gamma^{\rho}G^{\ast}_{R}(t)\gamma^{\mu}G_{R}(t-s_{1})\gamma^{\nu}\psi(s_{1})
\nn
&+&
A^{\nu}(s_{1})\bar{\psi}(s_{1})\gamma^{\nu}G^{\ast}_{R}(t-s_{1})\gamma^{\mu}G_{R}(t)\gamma^{\rho}\psi(0)
\label{eq:c1onfold}
\end{eqnarray}
The whole first order term will have $16$ terms. However, most contain a photon three point function and will vanish due to Furry's theorem. The nonvanishing terms will gives us
\begin{eqnarray}
C_{1}(t,\x)&=&-\frac{1}{4N_{F}}\int_{s_{i}s^{\prime}_{i}}\Big[
\mathcal{D}^{\nu\nu^{\prime}}_{W}(s_{1})(G^{\ast}_{W}(t)\gamma^{\mu}G_{R} (t-s_{1})\gamma^{\nu}G_{R}(s_{1})\gamma^{\rho}G_{W}(0)
\nn
&&
\gamma^{\mu^{\prime}}G^{\ast}_{R} (t-s^{\prime}_{1})\gamma^{\nu}G^{\ast}_{R}(s^{\prime}_{1})\gamma^{\rho}
\nn
&+&G^{\ast}_{W}(0)\gamma^{\rho}G^{\ast}_{R}(s_{1})\gamma^{\nu}G^{\ast}_{R}(t-s_{1})\gamma^{\mu}G_{W}(t)\gamma^{\rho^{\prime}}G^{\ast}_{R}(s^{\prime}_{1})\gamma^{\nu^{\prime}}G^{\ast}_{R}(t^{\prime}-s^{\prime}_{1})\gamma^{\mu^{\prime}}
\nn
&-&G^{\ast}_{W}(t)\gamma^{\mu}G_{R} (t-s_{1})\gamma^{\nu}G_{R}(s_{1})\gamma^{\rho}G^{\ast}_{W}(t)\gamma^{\rho^{\prime}}G^{\ast}_{R}(s^{\prime}_{1})\gamma^{\nu^{\prime}}G^{\ast}_{R}(t^{\prime}-s^{\prime}_{1})\gamma^{\mu^{\prime}}
\nn
&-&G_{W}(t)\gamma^{\rho}G^{\ast}_{R}(s_{1})\gamma^{\nu}G^{\ast}_{R}(t-s_{1})\gamma^{\mu}G_{W}(t)\gamma^{\mu^{\prime}}G^{\ast}_{R} (t-s^{\prime}_{1})\gamma^{\nu}G^{\ast}_{R}(s^{\prime}_{1})\gamma^{\rho})+...\Big]
\end{eqnarray}
where the primes will technically be contracted with the unprimed terms. These correspond to boxes with one rung with momenta flowing in opposite directions. The parts in between the Wightman functions are the same as the first order correction to the fermion ladder. These are pictured in Fig.~\ref{fig:currents}.
To second order
\begin{eqnarray}
[[[j_{\mu}(t),A^{\nu}j_{\nu}(s_{1})],A^{\rho}j_{\rho}(s_{2})],j_{\sigma}(0)]
&=&A^{\nu}(s_{1})A^{\rho}(s_{2})[[[j_{\mu}(t),j_{\nu}(s_{1})],j_{\rho}(s_{2})],j_{\sigma}(0)]
\nn
&+&[A^{\nu}(s_{1}),A^{\rho}(s_{2})][j_{\mu}(t),j_{\nu}(s_{1})][j_{\rho}(s_{2}),j_{\sigma}(0)]
\nn
&+&
[A^{\nu}(s_{1}),A^{\rho}(s_{2})][[j_{\mu}(t),\tilde{j}_{\nu}(s_{1})],j_{\sigma}(0)]j_{\rho}(s_{2})
\end{eqnarray}
By inspection, the first term will give us terms with two photon rails or a self energy correction. The second term will give the second type of rung appearing in the fermion diagram. This give four terms where the momentum flow in the boxes at either end can go in the opposite direction. The third term can give us a ladder with self energy corrections on the rails.

\section{Details of numerical computations}
\label{appendix:numerics}
Here we discuss the details of our numerical computation of $\Gamma_{\k}$ and $\lambda_{L}$. Our methods are similar to those used in \cite{Chowdhury:2017jzb}, with some adjustments made to address various issues. The most challenging aspect of the numerics is managing the light-cone singularities that appear throughout the computation, which require additional care. To compute the components of the polarization bubble, we dealt with this issue by a slight off-shift in the grid. From our analysis in App.~\ref{appendix:photonprop}, we know that all singularities that appear in the decay rate and the kernel should be integrable, so we can manually remove them during each integration.

The first step in our computation is to compute the analytic continuation of the $xx$ and $yy$ components of the polarization bubble. The imaginary parts are computed by numerical integration of Eqn.~\ref{eq:polarizationexpression} using a standard MATLAB routine. The integral is computed separately for $k^{0}<|\k|$ and $k^{0}>|\k|$ because of the non-analyticity at $k^{0}=|\k|$. The real parts are then obtained by a stable version of Kramers-Kronig given in the Eqn.~\ref{eq:kramerskronig}. These are then used to precompute the values of the spectral function, retarded propagator, and Wightman propagator on $200 \times 200$ grids for $k^{0}/T\in [0, 12.1]$, $|\k|/T \in [0, 12.21]$. These grids are then used to create two dimensional cubic interpolations for all propagators. $\Gamma_{\k}$ is calculated from Eqn.~\ref{eq:finalgamma} on a $60$ point grid for $|\k|/T \in [0, 12]$ where $60$ grid points are used for the angular integral. To calculate the exponent on varying grid sizes, we use a one dimensional spline interpolation for the decay rate.

The kernel is computed using Eqn.~\ref{eq:kernelcompute} on a $40\times40$ grid for $|\k|/T, |\k^{\prime}|/T \in [0, 12]$ where $40$ grid points are used for the angular integral. We find that most eigenvalues are complex with negative real parts, but the eigenvalue with the largest real part is completely real. This eigenvalue is identified as the Lyapunov exponent. There are also two eigenvalues which are exactly zero. We also compute the kernel using an interpolation on a grid that is slightly offset and find $\lambda_{L}\approx 4.06$.
\end{appendix}
\addcontentsline{toc}{section}{References}
\bibliographystyle{JHEP}
\bibliography{qed3scrambling}
\end{document}